\documentclass[aps,prd,onecolumn,groupedaddress,floatfix,longbibliography,superscriptaddress,notitlepage]{revtex4-1}

\usepackage{amsmath}
\usepackage{amssymb}
\usepackage{amsfonts}
\usepackage{graphicx}
\usepackage{bbold}
\usepackage{bm}
\usepackage{bm}
\usepackage{cancel}
\usepackage{times,float}
\usepackage{graphicx}
\usepackage[usenames,dvipsnames,svgnames]{xcolor}
\usepackage{hyperref}
\hypersetup{colorlinks=true, linkcolor=NavyBlue, citecolor=PineGreen,urlcolor=cyan}
\usepackage{multirow}
\usepackage{physics}

\begin{document}

\title{Radiation of a short linear antenna above a topologically insulating half-space}

\author{M. Ibarra-Meneses}
\email{martin\_i@ciencias.unam.mx}
\affiliation{Instituto de Ciencias Nucleares, Universidad Nacional Aut\'{o}noma de M\'{e}xico, 04510 Ciudad de M\'{e}xico, M\'{e}xico}

\author{A. Mart\'{i}n-Ruiz}
\email{alberto.martin@nucleares.unam.mx}
\affiliation{Instituto de Ciencias Nucleares, Universidad Nacional Aut\'{o}noma de M\'{e}xico, 04510 Ciudad de M\'{e}xico, M\'{e}xico}

\begin{abstract}
The topological magnetoelectric effect (TME) is  a unique macroscopic manifestation of quantum states of matter possessing topological order and it is described by axion electrodynamics. In three-dimensional topological insulators, for instance, the axion coupling is of the order of the fine structure constant, and hence a perturbative analysis of the field equations is plenty justified. In this paper we use Green's function techniques to obtain time-dependent solutions to the axion field equations in the presence of a planar domain-wall separating two media with different topological order. We apply our results to investigate the radiation of a short linear antenna near the domain-wall.  
\end{abstract}

\maketitle

\section{Introduction} \label{Intro}

Topological materials are not distinguished by local ordered parameters,  but by topological quantum numbers which could be described and computed using topological field theories in the low energy limit.  The best studied of these are the topological insulators (TIs), where the corresponding bulk band gap is similar to an ordinary insulator, but the boundary of the material hosts gapless states which are protected by the topological numbers of the bulk geometry \cite{RevModPhys.83.1057}. The experimental discovery of TIs in HgTe quantum wells \cite{doi:10.1126/science.1148047} and stoichiometric crystals \cite{doi:10.1126/science.1173034, 10.1038/nphys1274, PhysRevLett.105.136802} encourages further investigations regarding its exotic properties.

Apart from their spectroscopic distinguishing features, TIs have unconventional electromagnetic response described by axion electrodynamics, which is defined by the standard Maxwell Lagrangian plus the additional $\theta$-term (in SI units) $\mathcal{L} _{\theta} = (\alpha / \pi )  \sqrt{\epsilon _{0}/ \mu _{0}} \, \theta \, {\bf{E}} \cdot {\bf{B}}$, where ${\bf{E}}$ and ${\bf{B}}$ are the electromagnetic fields, $\alpha$ is the fine structure constant, and $\theta$ is an angular variable known in particle physics as the axion field or axion angle \cite{PhysRevLett.58.1799}. The axion field was originally proposed to solve the strong-CP problem in QCD \cite{PhysRevLett.38.1440, PhysRevLett.40.223, PhysRevLett.40.279}, but its weak interaction strength with standard model particles and an elegant production mechanism in the early universe make it a promising candidate to explain dark matter as well \cite{PRESKILL1983127, DINE1983137, ABBOTT1983133}. In the context of TIs, time-reversal (TR) symmetry indicates that $\theta = 0, \pi$ (mod $2\pi $), and hence the $\theta$-term has no effect on Maxwell equations in the bulk. The nontrivial topological property, a surface half-integer quantum Hall effect, manifests only at the boundary between two manifolds with different topologies, in which case $\theta$ is not globally constant. Therefore, from the macroscopic point of view, the axion angle $\theta$ can be viewed as an effective parameter characterizing the media, in the same footing as its permittivity $\epsilon$ and permeability $\mu$.

The $\theta$-term encodes the most salient feature of the electromagnetic response of TIs: the topological magnetoelectric effect, where an electric field can induce a magnetic polarization and a magnetic field can induce an electric polarization \cite{Fiebig_2005}. This is why, in the condensed matter literature, the axion angle is termed the topological magnetoelectric polarization (TMEP). A number of topological magnetoelectric effects have been predicted on the basis of this theory, namely, the appearance of charge and current densities of magnetic monopoles \cite{doi:10.1126/science.1167747, PhysRevLett.103.171601,PhysRevB.81.245125,  PhysRevD.92.125015, PhysRevA.100.042124,PhysRevD.94.085019, PhysRevD.98.056012, PhysRevX.9.011011}, the magneto-optical Faraday and Kerr rotation \cite{PhysRevB.80.113304, PhysRevB.84.205327, PhysRevA.94.033816, doi:10.1126/science.aaf5541, filipini2023new}, the Casimir effect \cite{PhysRevLett.106.020403, PhysRevB.84.045119, PhysRevLett.112.056804, PhysRevB.85.115102, PhysRevB.84.165409,PhysRevB.88.085421, Mart_n_Ruiz_2016}, as well as the influence of TIs on atomic systems nearby \cite{Song_2014,Fang_2015,PhysRevA.97.022502,Mart_n_Ruiz_2017,PhysRevA.102.013720,Castro_Enriquez_2020}.

Time-independent solutions to the axion field equations have been elaborated by using the electric-magnetic duality of $\mathbb{Z} _{2}$ TIs \cite{PhysRevLett.103.171601} as well as Green's function techniques \cite{PhysRevD.92.125015, PhysRevD.94.085019}. However, general solutions to the time-dependent case are still under investigation \cite{PhysRevD.99.116020, franca2021radiation, PhysRevB.105.155120}. In this work we aim to fill in this gap by computing, in a perturbative fashion, solutions to the time-dependent axion field equations in the presence of a planar domain wall separating two media with different topologies. The validity of a perturbative scheme for the solution of the field equations is as follows. In the high-energy context the validity is plenty justified: the axion-to-two- photon coupling $g _{\theta \gamma \gamma}$ which converts photons into axions is expected to be very small \cite{PhysRevD.99.055010}. In the condensed matter framework, the smallness of the $\theta$-term is controlled by the coupling $\alpha \theta$, which is small since $\alpha \approx 1/137$ and $\theta = \pi$ for a TI with a single Dirac fermion on its surface. In the hypothetical situation of a TI with a large number of Dirac fermions on its surface, our perturbative approach ceases to be valid as $\alpha \theta$ approaches unity. In view that exact solutions to the static axion field equations have been reported in the literature, we shall first validate our perturbative scheme by reproducing well-known results in this case. Next we will safely extend our analysis to the time-dependent case. Finally, we apply our results to study the radiation of a short linear antenna near the domain-wall.

The paper is organized as follows. In Sec. \ref{Theory} we briefly review the basics of axion electrodynamics. In Sec. \ref{Sec_Perturbative_Solutions} we solve, in a perturbative fashion, the static field equations and verify its consistency with previously reported results for the Green's function for two ponderable topological insulators separated by a planar interface. Section \ref{Time-dep-GF} is devoted to the evaluation of the time-dependent Green's function by assuming equal optical properties in the two media. This configuration allow us to isolate the topological effects. The details of technical calculations are relegated to the appendices. Finally, in Sec. \ref{Radiation_Antenna_Sec} we apply our result to investigate the radiation of a short linear antenna near a $\theta$ domain-wall.

\section{Topological field theory} \label{Theory}

The effective field theory governing the electromagnetic response of topological insulators, independently of the microscopic details, is defined by the action (in SI units) \cite{PhysRevB.78.195424}:
\begin{align}
S = \int d^{4}x \, \left[ \,  \frac{1}{2} \left(  \epsilon  {\bf{E}} ^{2} - \frac{1}{\mu}   {\bf{B}} ^{2} \right) + \frac{\alpha}{\pi} \sqrt{\frac{\epsilon _{0}}{\mu _{0}}} \, \theta \, {\bf{E}} \cdot {\bf{B}} \, \right] , \label{Action}
\end{align}
where $\alpha = e^{2}/ ( 4 \pi \epsilon _{0} \hbar c ) \simeq 1/137$ is the fine structure constant, $\epsilon$ and $\mu$ are the permittivity and permeability of the material, respectively, and $\theta$ is the topological magnetoelectric polarizability (axion field). Under periodic boundary conditions, the partition function and all physical quantities are invariant under shifts of $\theta$ by any multiple of $2 \pi$.  Since ${\bf{E}} \cdot {\bf{B}}$ is odd under TR, the only symmetry allowed values of $\theta$ are $0$ or $\pi$ (modulo $2\pi$).  This leads to the $\mathbb{Z}_{2}$ classification of three-dimensional TR invariant TIs. Clearly, the $\theta$-term in Eq. (\ref{Action}) has no effect on Maxwell equations in the bulk, since it is a total derivative term. The nontrivial topological property, a surface half-integer quantum Hall effect, manifests only when a TR-breaking perturbation is induced on the surface to gap the surface states, thereby converting the material into a full insulator. This can be achieved by introducing magnetic dopants to the surface \cite{doi:10.1126/science.1230905} or by the application of an external static magnetic field \cite{PhysRevLett.105.166803}. In this situation, $\theta$ is quantized in odd integer values of $\pi$, where the magnitude and sign of the multiple is controlled by  the strength and direction of the TR-breaking perturbation.

\begin{figure}
\includegraphics[scale=0.4]{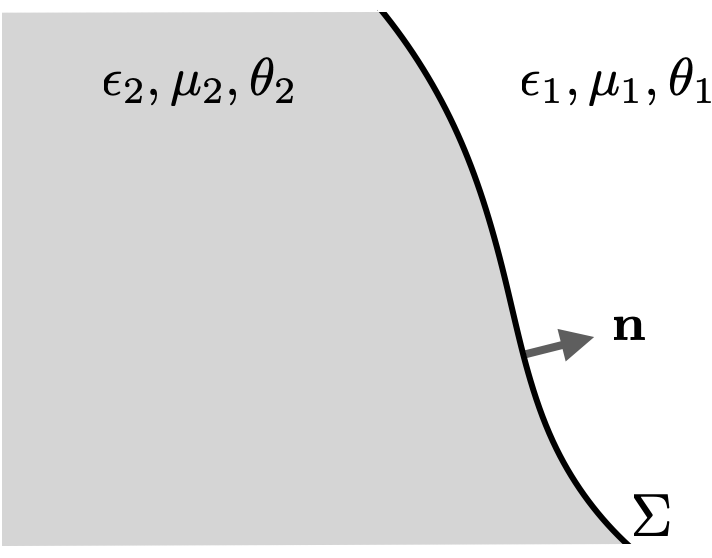}
\caption{Geometry of two semi-infinite topologically insulating media separated by the surface $\Sigma$. The media have different magnetoelectric polarizabilities, permittivities and permeabilities, as depicted.} \label{configuration}
\end{figure}

The electromagnetic response of a system in the presence of the axionic term is still described by the ordinary Maxwell equations, but the constituent relations which define the electric displacement ${\bf{D}}$ and the magnetic field ${\bf{H}}$ acquire an extra term proportional to $\theta$, i.e.  ${\bf{D}} = \epsilon {\bf{E}} + \alpha ( \theta / \pi ) \,  \sqrt{ \epsilon _{0} / \mu _{0}} {\bf{B}}$ and $ {\bf{H}} =\mu ^{-1} {\bf{B}} - \alpha ( \theta / \pi ) \,   \sqrt{ \epsilon _{0} / \mu _{0}}  \,  {\bf{E}} $. Here, the $\theta$-dependent terms, which arise from the axionic term in Eq. (\ref{Action}),  describe the topological magnetoelectric effect. The frequency components of the electromagnetic fields (introduced as usual: by making a Fourier decomposition in time of the sources and fields) obey the inhomogeneous equations
\begin{align}
\nabla \cdot \left[ \epsilon ({\bf{r}} , \omega ) \, {\bf{E}} ({\bf{r}} , \omega ) \right] = \rho ({\bf{r}} , \omega ) - \frac{\alpha}{\pi} \,  \sqrt{\frac{\epsilon _{0}}{\mu _{0}}}  \, \nabla \theta ({\bf{r}} ) \cdot {\bf{B}} ({\bf{r}} , \omega ), \label{GaussE} \\ \nabla \times \left[ \mu ^{-1} ({\bf{r}} , \omega ) \, {\bf{B}} ({\bf{r}} , \omega ) \right]  + i \omega \epsilon ({\bf{r}} , \omega ) {\bf{E}} ({\bf{r}} , \omega ) = {\bf{J}} ({\bf{r}} , \omega ) + \frac{\alpha}{\pi} \,  \sqrt{\frac{\epsilon _{0}}{\mu _{0}}}  \, \nabla \theta ({\bf{r}} ) \times {\bf{E}} ({\bf{r}} , \omega ) ,  \label{Ampere}
\end{align}
while the homogeneous equations are untouched due to gauge invariance of the action (\ref{Action}):
\begin{align}
\nabla \cdot {\bf{B}} ({\bf{r}} , \omega ) = 0 , \label{GaussB} \\[4pt] \nabla \times {\bf{E}} ({\bf{r}} , \omega ) = i \omega {\bf{B}} ({\bf{r}} , \omega ) . \label{Faraday}
\end{align}
One can further verify that the divergence equations (\ref{GaussE}) and (\ref{GaussB}) are subsumed within the curl equations (\ref{Ampere}) and (\ref{Faraday}), which can be verified by taking the divergence of the latters. We observe that equations (\ref{GaussE})-(\ref{Ampere}) depend only on the space gradient of the axion field. So, the $\theta$-term does not modify the field equations in the bulk of topological insulators. However, for a TI in contact with a conventional insulator, as shown in  Fig.  \ref{configuration}, the axion field is piecewise constant across the interface $\Sigma$ and then $\nabla \theta ({\bf{r}}) = ( \theta _{1} - \theta _{2} ) \delta (\Sigma )\hat{{\bf{n}}}$, where $\hat{{\bf{n}}}$ is the outward unit normal to $\Sigma$. These are the equations we aim to solve in a perturbative fashion along this paper. Our general approach will be to first Taylor expand the electromagnetic potentials in powers of the fine structure constant and then solve recursively the resulting equations. Before embarking to the solution of the time-dependent field equations, we first tackle the time-independent case and demonstrate consistency of our results with previously reported ones.

\section{Solution to the static axion field equations} \label{Sec_Perturbative_Solutions}

Let us consider the geometry shown in Fig.  \ref{configuration} with the interface $\Sigma$ being the plane $z=0$. The half-space $z<0$ is occupied by a topological medium with dielectric constant $\epsilon _{2}$, magnetic permeability $\mu _{2}$ and axion angle $\theta _{2}$, whereas the half-space $z>0$ is occupied by a medium with dielectric constant $\epsilon _{1}$, magnetic permeability $\mu _{1}$ and axion angle $\theta _{1}$. Then the inhomogeneity in the permittivity, permeability and the axion field are limited to a finite discontinuity across the planar surface $z=0$.

To proceed further we write the field equations (\ref{GaussE}) and (\ref{Ampere}) in terms of the electrostatic potential $\phi ({\bf{r}})$ and vector potential ${\bf{A}} ({\bf{r}})$. In the Coulomb gauge, the static electromagnetic potentials satisfy
\begin{align}
- \nabla \cdot \left[ \epsilon ({\bf{r}}  ) \, \nabla \phi ({\bf{r}}   ) \right] = \rho ({\bf{r}} ) - \tilde{\alpha} \,  \sqrt{\frac{\epsilon _{0}}{\mu _{0}}} \, \delta (z)  \, \hat{{\bf{e}}} _{z} \cdot \nabla \times {\bf{A}} ({\bf{r}} ), \label{GaussE-Static} \\ \nabla \times \left[ \mu ^{-1} ({\bf{r}} ) \, \nabla \times {\bf{A}} ({\bf{r}} ) \right] = {\bf{J}} ({\bf{r}} ) - \tilde{\alpha} \,  \sqrt{\frac{\epsilon _{0}}{\mu _{0}}} \, \delta (z)  \, \hat{{\bf{e}}} _{z} \times \nabla \phi ({\bf{r}} ) ,   \label{Ampere-Static}
\end{align}
where $\tilde{\alpha} = \alpha (\theta _{1} - \theta _{2}) / \pi$. To obtain a perturbative solution for the electromagnetic potentials in the presence of arbitrary external sources, we introduce the expansions $\phi ({\bf{r}}) = \phi ^{(0)} ({\bf{r}}) + \sum _{n=1} ^{\infty} \phi ^{(n)} ({\bf{r}})$ and ${\bf{A}} ({\bf{r}}) = {\bf{A}} ^{(0)} ({\bf{r}}) + \sum _{n=1} ^{\infty} {\bf{A}} ^{(n)} ({\bf{r}})$, where $\phi ^{(n)}$ and ${\bf{A}} ^{(n)} ({\bf{r}})$ are understood to vanish as $\tilde{\alpha} ^{n}$. We now substitute the power expansion into equations (\ref{GaussE-Static}) and (\ref{Ampere-Static}), and comparing coefficients of each power of $\tilde{\alpha}$ results in an infinite series of simultaneous equations. The zeroth-order equations are simply the Maxwell equations in the absence of the $\theta$-term,
\begin{align}
- \nabla \cdot \left[ \epsilon ({\bf{r}}  ) \, \nabla \phi ^{(0)} ({\bf{r}} ) \right] = \rho ({\bf{r}} )  ,  \qquad \qquad \nabla \times \left[ \mu ^{-1} ({\bf{r}} ) \, \nabla \times {\bf{A}} ^{(0)} ({\bf{r}} ) \right] = {\bf{J}} ({\bf{r}} ) . \label{Static-0-order} 
\end{align}
The $n$-order equations, which are the main concern in this work, are
\begin{align}
\nabla \cdot \left[ \epsilon ({\bf{r}}  ) \, \nabla \phi ^{(n)} ({\bf{r}}   ) \right] = \tilde{\alpha} \sqrt{\frac{\epsilon _{0}}{\mu _{0}}} \, \delta (z)  \, \hat{{\bf{e}}} _{z} \cdot \nabla \times {\bf{A}} ^{(n-1)} ({\bf{r}} ), \quad \nabla \times \left[ \mu ^{-1} ({\bf{r}} ) \, \nabla \times {\bf{A}} ^{(n)} ({\bf{r}} ) \right] = - \tilde{\alpha} \,  \sqrt{\frac{\epsilon _{0}}{\mu _{0}}} \, \delta (z)  \, \hat{{\bf{e}}} _{z} \times \nabla \phi ^{(n-1)} ({\bf{r}} ) .   \label{n-Order-Static}
\end{align}
The process of solving these equations begins with the solution to the zeroth-order equations (\ref{Static-0-order}), which involves the well-known electro- and magnetostatic Green's functions. We now adopt the following notation. Since translational invariance is broken in the $z$ direction due to the presence of the interface, it is natural decompose any vector ${\bf{C}}$ into components parallel and perpendicular to the normal direction $\hat{{\bf{e}}} _{z}$, i.e. ${\bf{C}} _{z} = C_{z}  \hat{{\bf{e}}} _{z}$ and ${\bf{C}} _{\perp} = C_{x}  \hat{{\bf{e}}} _{x} +  C_{y} \hat{{\bf{e}}} _{y} $, respectively. So, the general solution for the electrostatic potential is
\begin{align}
\phi ^{(0)} ({\bf{r}}) = \int G _{\epsilon} ({\bf{r}},{\bf{r}} ^{\prime}) \, \rho ({\bf{r}}^{\prime}) \, d^{3} {\bf{r}}^{\prime} , 
\end{align}
where the electric Green's function is given by
\begin{align}
G _{\epsilon} ({\bf{r}},{\bf{r}} ^{\prime}) = \frac{1}{4 \pi \epsilon _{1} } \left[ \frac{1}{ \sqrt{ \vert {\bf{r}} _{\perp } - {\bf{r}} ^{\prime} {} _{\!\! \perp } \vert ^{2} + ( z - z ^{\prime} ) ^{2}} } + \frac{\epsilon_1 - \epsilon_2}{\epsilon_1 + \epsilon_2} \frac{1}{ \sqrt{ \vert {\bf{r}} _{\perp } - {\bf{r}} ^{\prime} {} _{\!\! \perp } \vert ^{2} + (\vert z \vert + z') ^{2}} } \right] \label{Electric-GF}
\end{align}
in which ${\bf{r}} _{\perp } = x  \hat{{\bf{e}}} _{x} + y  \hat{{\bf{e}}} _{y}$ and ${\bf{r}} ^{\prime} {} _{\!\! \perp } = x ^{\prime}  \hat{{\bf{e}}} _{x} + y ^{\prime} \hat{{\bf{e}}} _{y}$ \cite{Schwinger}. The solution to the magnetostatic problem is a bit more subtle. Decomposing the current density and the vector potential into their components parallel and perpendicular to $\hat{{\bf{e}}} _{z}$, i.e. ${\bf{J}} = {\bf{J}} _{\perp} + {\bf{J}} _{z}$ and ${\bf{A}} = {\bf{A}} _{\perp} + {\bf{A}} _{z}$, respectively, the general solution for the vector potential, in the half-space $z>0$, can be written as
\begin{align}
{\bf{A}} ^{(0)} _{\xi} ({\bf{r}}) = \int G _{\mu} ^{(\xi )} ({\bf{r}},{\bf{r}} ^{\prime}) \, {\bf{J}} _{\xi} ({\bf{r}}^{\prime}) \, d^{3} {\bf{r}}^{\prime} ,  \qquad \xi = \, \perp , z
\end{align}
while for the region $z<0$ the solution is simply ${\bf{A}} ^{(0)} ({\bf{r}}) = \int G _{\mu} ^{(\perp )} ({\bf{r}},{\bf{r}} ^{\prime}) \, {\bf{J}}  ({\bf{r}}^{\prime}) \, d^{3} {\bf{r}}^{\prime}$. The corresponding magnetic Green's function is given by \cite{Schwinger}
\begin{align}
    G _{\mu} ^{(\xi )} ({\bf{r}},{\bf{r}} ^{\prime}) &=  \frac{\mu _{1}}{4 \pi } \left[ \frac{1}{ \sqrt{ \vert {\bf{r}} _{\perp } - {\bf{r}} ^{\prime} {} _{\!\! \perp } \vert ^{2} + ( z - z ^{\prime} ) ^{2}} } + \chi \, \frac{ \mu _{2} - \mu _{1} }{ \mu _{2} + \mu _{1} } \frac{1}{ \sqrt{ \vert {\bf{r}} _{\perp } - {\bf{r}} ^{\prime} {} _{\!\! \perp } \vert ^{2} + (\vert z \vert + z ^{\prime}) ^{2}} } \right] ,  \label{xi-Magnetic-GF}
\end{align}
where $\chi = + 1 \, (-1)$ for $\xi = \, \perp (z)$.

With all of the above we are ready to solve equations (\ref{n-Order-Static}). This is a simple but not straightforward task, so we relegate the details to Appendix \ref{Detailed-calculations}. The final results, up to second order in the parameter $\tilde{\alpha}$, are
\begin{align}
    \phi ({\bf{r}}) = \phi ^{(0)} ({\bf{r}}) + \int \left[ G _{0i} ^{(1)} ({\bf{r}},{\bf{r}} ^{\prime}) \, J_{i} ({\bf{r}} ^{\prime}) + G _{00} ^{(2)} ({\bf{r}},{\bf{r}} ^{\prime}) \, \rho ({\bf{r}} ^{\prime})  \right] d ^{3} {\bf{r}} ^{\prime} , \label{Phi-Main} \\ A _{i} ({\bf{r}}) = A _{i} ^{(0)} ({\bf{r}}) + \int \left[  G _{0i} ^{(1)} ({\bf{r}},{\bf{r}} ^{\prime}) \, \rho ({\bf{r}} ^{\prime}) + G _{ij} ^{(2)} ({\bf{r}},{\bf{r}} ^{\prime}) \, J_{j} ({\bf{r}} ^{\prime}) \right] d ^{3} {\bf{r}} ^{\prime} ,  \label{A-Main}
\end{align}
where the indices $i,j$ range over the transverse coordinates $x,y$, and
\begin{align}
    G _{0i} ^{(1)} ({\bf{r}},{\bf{r}} ^{\prime}) &= \frac{\tilde{\alpha} \sqrt{\epsilon _{0} / \mu _{0}} }{2 \pi ( \epsilon _{1} + \epsilon _{2} )( 1/ \mu _{1} + 1 / \mu _{2} )}  \, \frac{ \epsilon _{zji}  \, ( {\bf{r}} _{\perp} - {\bf{r}} _{\perp} ^{\prime } ) _{j} }{\vert {\bf{r}} _{\perp} - {\bf{r}} _{\perp} ^{\prime } \vert ^{2}} \, \left[ 1 - \frac{\vert z  \vert + z ^{\prime} }{ \sqrt{ \vert {\bf{r}} _{\perp} - {\bf{r}} _{\perp} ^{\prime } \vert ^{2} + ( \vert z  \vert + z ^{\prime} ) ^{2} }} \right] ,  \label{IntG0i(1)-Main} \\ G _{00} ^{(2)} ({\bf{r}},{\bf{r}} ^{\prime}) &= \frac{( \tilde{\alpha} \sqrt{\epsilon _{0} / \mu _{0}} \, ) ^{2}}{2 \pi ( \epsilon _{1} + \epsilon _{2} ) ^{2} (1 / \mu _{1} + 1 / \mu _{2})  } \, \frac{1}{ \sqrt{ \vert {\bf{r}} _{\perp } - {\bf{r}} ^{\prime} {} _{\!\! \perp } \vert ^{2} + (\vert z \vert  + z ^{\prime}) ^{2} } }  , \label{IntG00(2)-Main} \\ G _{ij} ^{(2)} ({\bf{r}},{\bf{r}} ^{\prime}) &= \frac{( \tilde{\alpha} \sqrt{\epsilon _{0} / \mu _{0}} \, ) ^{2}}{2 \pi ( \epsilon _{1} + \epsilon _{2} ) (1 / \mu _{1} + 1 / \mu _{2}) ^{2} }   \; \left[   \frac{\delta _{ij}}{ \sqrt{ \vert {\bf{r}} _{\perp } - {\bf{r}} ^{\prime} {} _{\!\! \perp } \vert ^{2} + (\vert z \vert  + z ^{\prime}) ^{2} } } + \partial _{i} K _{j} ({\bf{r}},{\bf{r}} ^{\prime}) \right] , \label{IntGij(2)-Main}
\end{align}
with the definition
\begin{align}
    K _{j} ({\bf{r}},{\bf{r}} ^{\prime}) = \frac{({\bf{r}} _{\perp }  - {\bf{r}} _{\perp } ^{\prime}) _{j}}{\vert {\bf{r}} _{\perp }  - {\bf{r}} _{\perp } ^{\prime} \vert ^{2} } \, \left( \sqrt{\vert {\bf{r}} _{\perp }  - {\bf{r}} _{\perp } ^{\prime} \vert ^{2} + (\vert z \vert  + z ^{\prime}) ^{2} } - (\vert z \vert  + z ^{\prime}) \right) .
\end{align}
Clearly, the electric (vector) potential is even (odd) under time-reversal operation, as expected.

\subsection*{Image magnetic monopole effect}

As a consistency check of our results, we shall verify the image magnetic monopole effect, which has been extensively discussed in the literature by using different approaches, for example, the method of images \cite{doi:10.1126/science.1167747}, the electric-magnetic duality of TIs \cite{PhysRevLett.103.171601} and Green's function techniques \cite{PhysRevD.94.085019, PhysRevD.98.056012}. This elusive effect  consists in the appearance of an image magnetic monopole when an electric charge is placed at a certain distance from the TI's surface. 

We now consider for the charge density a pointlike electric charge $q$ at $z=z_{0}$. Equations (\ref{Phi-Main}) and (\ref{A-Main}), together with the Green's functions (\ref{IntG0i(1)-Main})-(\ref{IntGij(2)-Main}), give the electric and vector potentials. For $z>0$, the electric potential can be interpreted as due to two pointlike electric charges, one of strength $q$ at $z=z_{0}$, and the other, the image charge, of strength
\begin{align}
    q ^{\prime} = q \left[  \frac{\epsilon_1 - \epsilon_2}{\epsilon_1 + \epsilon_2} +  \frac{2 \epsilon _{1} ( \tilde{\alpha} \sqrt{\epsilon _{0} / \mu _{0}} \, ) ^{2}}{( \epsilon _{1} + \epsilon _{2} ) ^{2} (1 / \mu _{1} + 1 / \mu _{2})  }  \right] \label{ImageCharge}
\end{align}
at $-z _{0}$. For $z<0$ only one pointlike electric charge appears, of strength $q ^{\prime} + q$ located at $z_{0}$. Unremarkably, the above expression reduce to the standard textbook one in the nontopological limit ($\tilde{\alpha} = 0$) \cite{Schwinger}. Due to the magnetoelectricy of TIs, a magnetic field is induced. The corresponding vector potential is found to be
\begin{align}
    {\bf{A}} ({\bf{r}}) = \frac{g}{4 \pi}  \, \frac{ x \hat{{\bf{e}}} _{y} - y \hat{{\bf{e}}} _{x} }{r _{\perp} ^{2}} \, \left[ 1 - \frac{\vert z  \vert + z _{0} }{ \sqrt{ r _{\perp} ^{2} + ( \vert z  \vert + z _{0} ) ^{2} }} \right] ,
\end{align}
which we identify as the vector potential of a straight vortex line or Dirac string over the $z$ axis. For $z>0$ ($z<0$) the string is over the negative (positive) $z$ axis ending at a magnetic monopole at $- z _{0}$ ($z _{0}$). In other words, the magnetic field for $z>0$ can be interpreted as that of a magnetic monopole of strength
\begin{align}
    g =  \frac{2q \tilde{\alpha} \sqrt{\epsilon _{0} / \mu _{0}} }{( \epsilon _{1} + \epsilon _{2} )( 1/ \mu _{1} + 1 / \mu _{2} )} \label{MagneticCharge} 
\end{align}
at the image point $-z_{0}$. Similarly, for $z<0$ the magnetic field can be understood as originated by a monopole of strength $-g$ located at $z_{0}$. Our results are in a agreement with the ones reported in Refs. \cite{doi:10.1126/science.1167747, PhysRevLett.103.171601, PhysRevD.94.085019, PhysRevD.98.056012}. This can be further verified by power expanding the image electric and magnetic charges up to order $\tilde{\alpha} ^{2}$, thus yielding our results (\ref{ImageCharge}) and (\ref{MagneticCharge}), respectively.

\section{Time-dependent solutions to the axion field equations} \label{Time-dep-GF}

In this Section we consider the same configuration as in Sec. \ref{Sec_Perturbative_Solutions}, but we shall investigate now time-dependent solutions to the field equations (\ref{GaussE})-(\ref{Faraday}).  In order to isolate the contribution arising from the surface Hall effect, we restrict ourselves to the case in which the permittivities are the same, i.e. $\epsilon _{1} = \epsilon _{2} \equiv \epsilon$, but the topological magnetoelectric polarizability is a piecewise constant function: $\theta _{1}$ for $z>0$ and $\theta _{2}$ for $z<0$. We assume the sources located in the region $z>0$.

We now introduce the field-potential relationships in the frequency domain, ${\bf{B}} ({\bf{r}} , \omega ) = \nabla \times {\bf{A}} ({\bf{r}} , \omega )$ and ${\bf{E}} ({\bf{r}} , \omega ) = - \nabla \phi ({\bf{r}} , \omega ) + i \omega {\bf{A}} ({\bf{r}} , \omega )$. In the Lorentz gauge,  $\nabla \cdot {\bf{A}} ({\bf{r}} , \omega ) = i \frac{\omega}{v ^{2}} \phi ({\bf{r}} , \omega )$ with $v = 1 / \sqrt{\epsilon \mu}$, the electromagnetic potentials satisfy
\begin{align}
- \left( \nabla ^{2} + k _{0} ^{2} \right) \phi ({\bf{r}}  , \omega )  &= \frac{\rho ({\bf{r}}  , \omega ) }{\epsilon }  - \frac{ \tilde{\alpha} }{\epsilon } \sqrt{\frac{\epsilon _{0}}{\mu _{0}}} \, \delta (z)  \, \hat{{\bf{e}}} _{z} \cdot \nabla \times {\bf{A}} ({\bf{r}}  , \omega ) , \label{PotPhi-Theta} \\ - \left( \nabla ^{2} + k _{0} ^{2} \right) {\bf{A}} ({\bf{r}}  , \omega ) &= \mu \,  {\bf{J}} ({\bf{r}}  , \omega )  - \tilde{\alpha} \mu \sqrt{\frac{\epsilon _{0}}{\mu _{0}}} \, \delta (z)  \, \hat{{\bf{e}}} _{z}  \times \left[ \nabla \phi ({\bf{r}}  , \omega )  - i \omega {\bf{A}} ({\bf{r}}  , \omega ) \right] , \label{PotA-Theta}
\end{align}
where $k _{0} = \omega / v$.  
  To solve this system of equations we follow the same perturbative program than before: we introduce the expansions $\phi ({\bf{r}}  , \omega ) = \phi ^{(0)} ({\bf{r}}  , \omega ) + \sum _{n=1} ^{\infty} \phi ^{(n)} ({\bf{r}}  , \omega ) $ and ${\bf{A}} ({\bf{r}}  , \omega ) = {\bf{A}} ^{(0)}  ({\bf{r}}  , \omega ) + \sum _{n=1} ^{\infty} {\bf{A}} ^{(n)} ({\bf{r}}  , \omega )$, where $\phi ^{(n)} ({\bf{r}} , \omega )$ and ${\bf{A}} ^{(n)} ({\bf{r}}  , \omega )$ vanish as $\tilde{\alpha} ^{n} $, and solve the system recursively. The zeroth-order equations are exactly Maxwell equations for $\tilde{\alpha} = 0$, i.e.
 \begin{align}
- \left( \nabla ^{2} + k _{0} ^{2} \right) \phi ^{(0)}  ({\bf{r}} , \omega) &= \frac{\rho ({\bf{r}} , \omega) }{\epsilon  } , \label{PotPhi-Theta0} \\ - \left( \nabla ^{2} + k _{0} ^{2} \right) {\bf{A}} ^{(0)} ({\bf{r}} , \omega)  &= \mu  \,  {\bf{J}} ({\bf{r}} , \omega)  , \label{PotA-Theta0}
\end{align}
while the $n$-order potentials are determined by
\begin{align}
- \left( \nabla ^{2} + k _{0} ^{2} \right) \phi ^{(n)} ({\bf{r}} , \omega) &= - \frac{ \tilde{\alpha} }{\epsilon } \sqrt{\frac{\epsilon _{0}}{\mu _{0}}} \, \delta (z)  \, \hat{{\bf{e}}} _{z} \cdot \nabla \times {\bf{A}} ^{(n-1)} ({\bf{r}} , \omega ) , \label{PotPhi-ThetaN} \\ - \left( \nabla ^{2} + k _{0} ^{2} \right) {\bf{A}} ^{(n)} ({\bf{r}}  , \omega ) &=  - \tilde{\alpha} \mu \sqrt{\frac{\epsilon _{0}}{\mu _{0}}} \, \delta (z)  \, \hat{{\bf{e}}} _{z}  \times \left[ \nabla \phi ^{(n-1)} ({\bf{r}}  , \omega )  - i \omega {\bf{A}} ^{(n-1)} ({\bf{r}}  , \omega ) \right] . \label{PotA-ThetaN}
\end{align}
The first step towards solving the higher-order equations (\ref{PotPhi-ThetaN})-(\ref{PotA-ThetaN}) is to solve the zeroth-order equations (\ref{PotPhi-Theta0})-(\ref{PotA-Theta0}). This requires the well-known frequency-dependent Green's function of electromagnetism \cite{Schwinger}:
\begin{align}
G _{\omega} ({\bf{r}},{\bf{r}}') =  \frac{e ^{i k _{0} \vert {\bf{r}} - {\bf{r}}' \vert  }}{ 4 \pi \vert {\bf{r}} - {\bf{r}}' \vert } , \label{Usual_Green_Frequency}
\end{align} 
such that
\begin{align}
\phi ^{(0)}  ({\bf{r}} , \omega)  = \frac{1}{\epsilon} \int G _{\omega} ({\bf{r}},{\bf{r}}') \, \rho ({\bf{r}}' , \omega) \, d ^{3} {\bf{r}}' , \qquad  {\bf{A}} ^{(0)}  ({\bf{r}} , \omega)  = \mu \int G _{\omega} ({\bf{r}},{\bf{r}}') \,  {\bf{J}} ({\bf{r}}' , \omega) \, d ^{3} {\bf{r}}' . \label{Zeroth-order-T_dep}
\end{align}
With these results we can evaluate now equations (\ref{PotPhi-ThetaN}) and (\ref{PotA-ThetaN}).  In this case we solve the system up to first order in the parameter $\tilde{\alpha}$. As shown in the Appendix \ref{Derivation_GF_omega}, for the vector potential one gets:
\begin{align}
A _{i} ({\bf{r}} , \omega) = A _{i} ^{(0)} ({\bf{r}} , \omega) + \int   G _{ij} ^{(1)} ({\bf{r}},{\bf{r}} ^{\prime}) \, J_{j} ({\bf{r}} ^{\prime} , \omega) \, d ^{3} {\bf{r}} ^{\prime} ,  \label{A-Main-time}
\end{align}
and for the scalar potential one spot the identity $i \frac{\omega}{v ^{2}} \phi ({\bf{r}} , \omega) = \nabla \cdot {\bf{A}} ({\bf{r}} , \omega)$.  Here,  
\begin{align}
G _{ij} ^{(1)} ({\bf{r}},{\bf{r}} ^{\prime}) &=  i \frac{1}{\omega} \left[ (\hat{\bf{e}} _{z} \times \nabla ) _{i} \partial _{j}  - k _{0} ^{2} \,  \epsilon _{ijz}   \right]   \mathcal{G} _{\omega}  ({\bf{r}} , {\bf{r}}' ) , \label{Gij-time}
\end{align}
where
\begin{align}
\mathcal{G} _{\omega} ({\bf{r}} , {\bf{r}}' ) = - \frac{\tilde{\alpha}}{8 \pi} \frac{ \mu}{\epsilon}  \sqrt{\frac{\epsilon _{0}}{\mu _{0}}} \int _{0} ^{\infty} d k _{\perp}  \frac{e ^{i \sqrt{ k _{0} ^{2}  - k _{\perp} ^{2} } ( \vert z \vert + \vert z ^{\prime} \vert )}}{ k _{0} ^{2} - k _{\perp} ^{2} }  k _{\perp} \, J _{0} ( k _{\perp} \vert {\bf{r}} _{\perp} - {\bf{r}} ^{\prime} _{\perp} \vert ) .  \label{ExactG_omega}
\end{align}
A detailed derivation of this expression can be found in the Appendix \ref{ExactApp}.  It is worth to mention that the exact electromagnetic potentials (and hence the fields) must be calculated using Eq. (\ref{ExactG_omega}), which unfortunately cannot be evaluated in an analytical fashion.  However, for many practical calculations, one can obtain an approximated expression valid for $\vert {\bf{r}} _{\perp} - {\bf{r}} ^{\prime} _{\perp} \vert \gg 1 $ and $ \vert z \vert + \vert z ^{\prime} \vert  \gg 1 $. In the Appendix \ref{Far_field_App} we derive, by two different methods,  the aforementioned expression.  The approximation relays in the fact that the integrand in Eq. (\ref{ExactG_omega}) is a rapidly oscillating function for large $\vert {\bf{r}} _{\perp} - {\bf{r}} ^{\prime} _{\perp} \vert $ and $ \vert z \vert + \vert z ^{\prime} \vert $, and hence the leading contribution can be obtained by using, for example, the stationary phase approximation. The final result is
\begin{align}
\mathcal{G} _{\omega} ({\bf{r}} , {\bf{r}}' ) \approx    i  \frac{\tilde{\alpha}}{8 \pi} \frac{ \mu}{\epsilon}  \sqrt{\frac{\epsilon _{0}}{\mu _{0}}} \frac{e ^{ i k _{0} \sqrt{ \vert {\bf{r}} _{\perp} - {\bf{r}} ^{\prime} _{\perp} \vert ^{2} + ( \vert z \vert + \vert z ^{\prime} \vert ) ^{2} } }   }{ k _{0} ( \vert z \vert + \vert z ^{\prime} \vert  ) }  .  \label{Aprox_G_omega}
\end{align}
Once we have found the indexed Green's function, we are ready to tackle problems involving different time-dependent sources near the domain wall.

\section{Radiation of a short linear antenna near a $\theta$ domain-wall} \label{Radiation_Antenna_Sec}

The problem of a short linear antenna near a dielectric half-space has attracted great attention for over the years due to its importance to many practical applications, in particular, in the context of THz applications \cite{10.1063/1.2171488}, near-field optics \cite{Novotny:97}, plasmonics \cite{10.1063/1.1951057} and nanophotonics \cite{10.1109/MNANO.2008.931481}. The exact solution to this problem was given first by Sommerfeld in 1909 \cite{https://doi.org/10.1002/andp.19093330402}. Since then numerous investigations have been carried out in this subject. The main goal of this Section is to address the problem of the radiation pattern produced by a short linear antenna placed near the boundary between two $\theta$-half spaces. To this end, we shall employ the theory developed in the previous Section.

\begin{figure}
\includegraphics[scale=0.4]{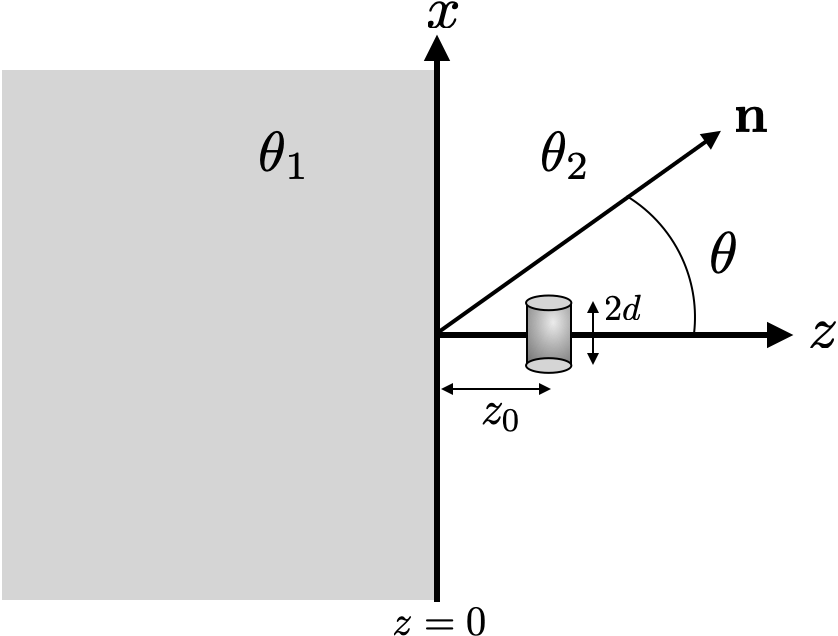}
\caption{Short linear antenna (of length $2d$ and frequency $\omega$) at a distance $z_{0}$ from the planar interface between two topologically insulating media.} \label{antenna_config}
\end{figure}

The geometry of the problem is shown in Fig.  \ref{antenna_config}. The left half-space $z<0$ is occupied by a topological insulator with axion angle $\theta _{1}$, while the right half-space $z>0$ is occupied by a TI with axion angle $\theta _{2}$. We consider a simple model for a transmitting antenna oriented parallel to the interface, say along the the $x$-axis and placed  at a distance $z_{0}$ from the interface, with uniform current flowing along a conductor of length $2d$. We assume that this current varies sinusoidally with time as
\begin{align}
{\bf{J}} ({\bf{r}} , t ) = I _{0}  \sin (\omega _{0} t) \,  \delta (y) \delta (z-z _{0}) \Theta (d - \vert x \vert ) \, \hat{{\bf{e}}}_{x}  , \label{Current-Antenna}
\end{align}
where $I _{0}$ and $\omega _{0}$ denotes the current amplitude  and angular frequency, respectively, the $\delta$'s are Dirac deltas and $\Theta$ is the Heaviside step function. Because the current is discontinuous at the ends, charge must be deposited there, being of opposite sign at each end, thus forming an electric dipole. The exact zeroth-order scalar and vector potentials can be calculated in a simple fashion and can be found in a variety of textbooks on electrodynamics \cite{Schwinger}. The only nonzero component of the vector potential (which is in the same direction as the current) in the time-domain is
\begin{align}
A _{x} ^{(0)}  ({\bf{r}} ,  t ) = \frac{\mu  I _{0} }{4 \pi } \int _{-d} ^{+d} dx'  \frac{1}{ \vert {\bf{r}} - {\bf{r}}'  \vert } \,  \sin \left[ \omega _{0} \left( t -  \frac{\vert {\bf{r}} - {\bf{r}}'   \vert }{c} \right) \right]  , 
\end{align}
where ${\bf{r}} ' (x') = x' \hat{{\bf{e}}}_{x} + z _{0} \hat{{\bf{e}}}_{z} $.  The result for the scalar potential in the time-domain is
\begin{align}
 \phi  ^{(0)}  ({\bf{r}} ,  t )  =  - \frac{ I _{0} }{4 \pi \epsilon \omega _{0} } \left[ \frac{\cos \left[ \omega _{0} \left( t -  \frac{\vert {\bf{r}} - {\bf{r}} _{+} \vert }{c} \right) \right] }{ \vert {\bf{r}} - {\bf{r}} _{+} \vert } - \frac{\cos \left[ \omega _{0} \left( t -  \frac{\vert {\bf{r}} - {\bf{r}} _{-} \vert }{c} \right) \right] }{ \vert {\bf{r}} - {\bf{r}} _{-} \vert } \right]   , 
\end{align}
where ${\bf{r}} _{\pm} = {\bf{r}} ' ( \pm d) $.  One can easily verify that these potentials satisfy the Lorentz gauge condition. 

\begin{figure}
\includegraphics[scale=0.54]{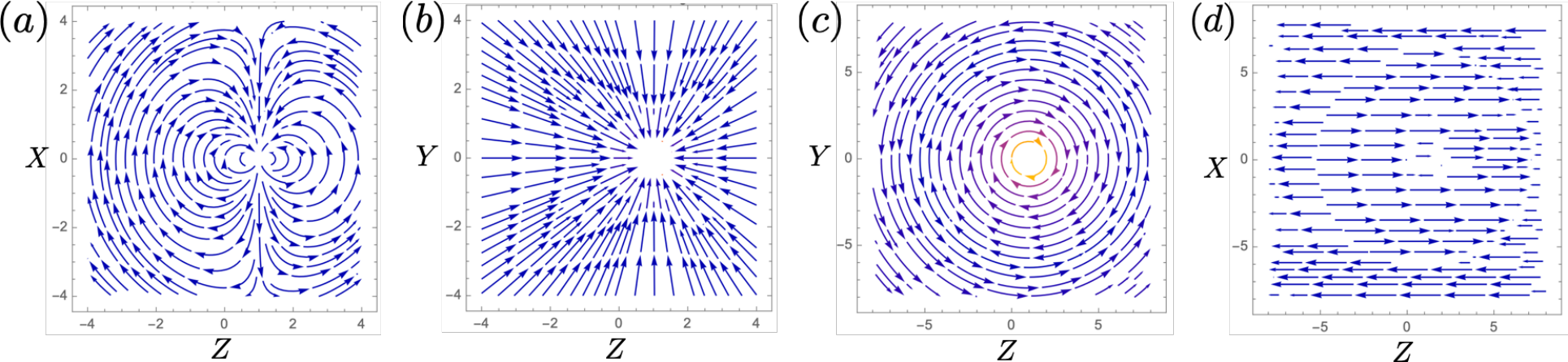}
\caption{Illustration of cross-sections of the normalized electric field [panels (a) and (b)] and magnetic field [panels (c) and (d)] generated by a short linear antenna in vacuum at $z _{0}$. } \label{E_Field_NoTop}
\end{figure}

\begin{figure}
\includegraphics[scale=0.54]{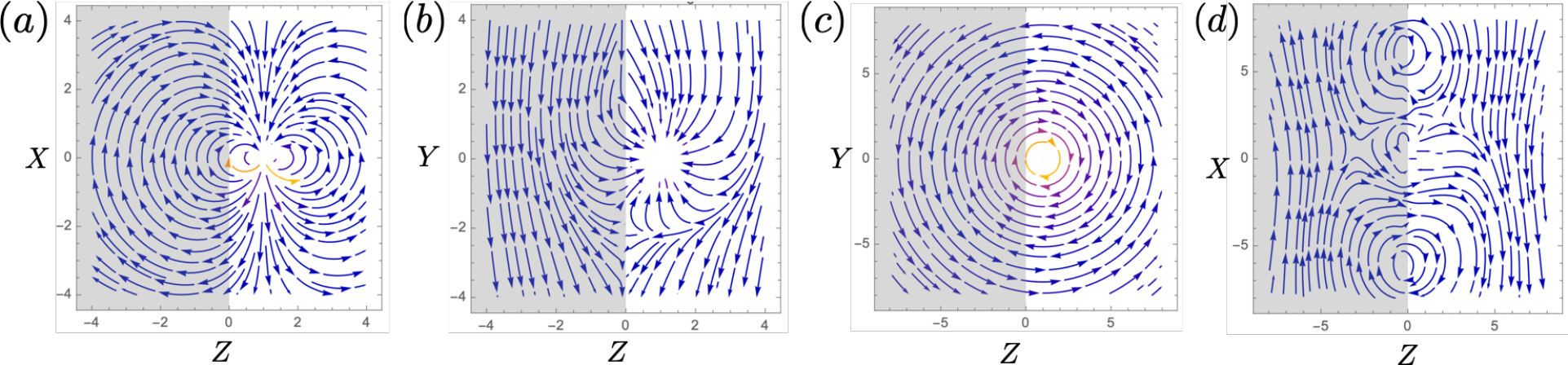}
\caption{Illustration of cross-sections of the normalized electric field [panels (a) and (b)] and magnetic field [panels (c) and (d)] generated by a short linear antenna at a distance $z _{0}$ from a $\theta$-domain wall at $z=0$. } \label{B_Field_NoTop}
\end{figure}

In order to evince the impact of the axion field, we first evaluate numerically the electromagnetic potentials and fields at first order in $\theta$ by using Eq. (\ref{A-Main-time}) together with the exact Green's function (\ref{Gij-time}) with $\mathcal{G} _{\omega}$ given by Eq. (\ref{ExactG_omega}).  To this end we introduce the dimensionless quantities $X= x/z _{0}$, $Y= y/z _{0}$ and $Z= z/z _{0}$.  Figure \ref{E_Field_NoTop} shows the electric and magnetic fields in the case $\theta = 0$, i.e. in the absence of the topological magnetoelectricity.  Panels (a) and (b) show the electric field lines  (in units of $E _{0} = \frac{\mu  I v}{4 \pi z _{0}}$) on the planes $Y=0$ and $X=d/Z_{0}$ (where the end-points of the antenna are located) for $\omega t = \pi /4$, respectively.  They exhibit the expected behavior: dipole-like in panel (a) and almost point-like in panel (b).Panels (c) and (d) show the magnetic field lines  (in units of $B _{0} = E _{0}/v$) on the planes $X=d/z_{0}$ and $Y=0$  for $\omega t = \pi /4$, respectively.  The former exhibits the usual current-encircling behavior dictated by the Ampère law, while the latter exhibits the expected results, whose filed lines invert their direction periodically. In order to show the full electromagnetic fields, including the effects of the magnetoelectric coupling, we have to introduce further dimensionless quantities besides the coordinates $(X,Y,Z)$, namely, $T = t/ \tau$, $D = d/z _{0}$, $\Lambda = \omega _{0} z _{0} / c$ and $\tau = z _{0} / c$.  This is so because the appearance of an additional length scale in the system: the antenna-surface distance $z_{0}$.  In Fig. \ref{B_Field_NoTop} we show the total electromagnetic fields for $\theta = \pi$, which is a typical value for topological insulators.  We observe that the cross-sections $X=0$ exhibit the same behavior as in the case $\theta = 0$, for both, the electric (panel (a)) and magnetic (panel (c)) fields; however, as we move from $X=0$ the field lines deviates from the nontopological case.  Panels (b) and (d) show the $Y=0$ cross-section, in which case we observe a considerably different behavior, which cannot be interpreted in a simple manner.  Indeed, the charge and current densities sourcing these fields are far from trivial, but they can be understood in terms of the half-integer Hall current at the interface.  The normalized Hall current ${\bf{j}} _{\mbox{\scriptsize Hall } } $  (in units of $\sigma _{\mbox{\scriptsize Hall } } E _{0}$, with $\sigma _{\mbox{\scriptsize Hall } } = e^{2}/2h$) is
\begin{align}
{\bf{j}} _{\mbox{\scriptsize Hall } } ({\bf{r}} _{\perp} ,  t ) = \hat{{\bf{e}}} _{x} \partial _{Y} F (X,Y)  + \hat{{\bf{e}}} _{y} \left[  \partial _{X} F (X,Y) -  \Lambda ^{2} G (X,Y)  \right] ,
\end{align}
where
\begin{align}
F (X,Y) =  \frac{\cos \left[ \Lambda (  T -  R _{+} )  \right] }{ R _{+} } - \frac{\cos \left[ \Lambda ( T -  R _{-} )  \right] }{ R _{-}  }  , \qquad G (X,Y) = \int _{-D} ^{+D} dX'  \frac{ \cos \left[ \Lambda ( T-S (X') ) \right] }{ S (X')} ,
\end{align}
with $S (X') = \sqrt{ (X-X')^{2} + Y^{2}+1 }$ and $R _{\pm} =  \sqrt{ (X \mp D ) ^{2} + Y ^{2} + 1 } $. 

Now we are interested in the fields that survive at large distances from the source, i.e. in the radiation zone, which requires approximating the potentials for $r \gg \lambda $ and $r \gg d$, where $\lambda = 2 \pi c/ \omega $ is the wavelength. To this end we use the expansions $ \vert {\bf{r}} - {\bf{r}}'   \vert \approx r - {\bf{n}} \cdot {\bf{r}}'$ and $ \vert {\bf{r}} - {\bf{r}} _{\pm}   \vert \approx r - \frac{zz_{0}}{r} \mp \frac{xd}{r}$, where ${\bf{n}} = {\bf{r}} / r$. All in all, the zeroth order potentials (up to order $1/r$) become
\begin{align}
 A _{x} ^{(0)}  ({\bf{r}} ,  t )  \approx   \frac{\mu  I _{0} }{2 \pi }  \sin \left[ \omega _{0} \left( t -  \frac{ r ^{2} - z _{0} z }{v r}  \right) \right]   \sin  \left( \frac{ \omega _{0} d    }{v}  \frac{x}{r} \right)     \frac{v}{\omega _{0} x }  , \qquad \phi  ^{(0)}  ({\bf{r}} ,  t ) =   v \frac{x}{r}  \, A _{x} ^{(0)}  ({\bf{r}} ,  t )     .
\end{align}
Using these potentials we find the electromagnetic fields in the radiation zone:
\begin{align}
{\bf{B}} ^{(0)} =  \frac{\mu  I _{0} }{2 \pi } \,   \cos \left[ \omega _{0} \left( t -  \frac{ r ^{2} - z _{0} z }{v r}  \right) \right]   \sin  \left( \frac{ \omega _{0} d    }{v}  \frac{x}{r} \right)   \, \frac{1}{x} \hat{{\bf{e}}} _{x} \times \hat{{\bf{n}}}  , \qquad  {\bf{E}} ^{(0)}  =  - v \,  \hat{{\bf{n}}}  \times   {\bf{B}} ^{(0)}  . \label{B0}
\end{align}
Following the same program one can derive expressions for the electromagnetic potentials at first-order. The details are relegated to the Appendix \ref{App_Antenna}. Using Eq. (\ref{A-Main-time}) one finds that the nonzero components of the vector potential in the time-domain at first-order become
\begin{align}
A _{x}  ^{(1)} ({\bf{r}} , t ) &=  \frac{\tilde{\alpha}}{8 \pi } \frac{\mu}{\epsilon} \sqrt{\frac{\epsilon _{0}}{\mu _{0}}} \frac{I _{0} }{ \omega _{0} k _{0} (\vert z \vert + z _{0})}  \frac{\partial ^{2}}{\partial x \partial y}   \int _{-d} ^{+d}  d x ^{\prime}  \sin  \left[  \omega _{0} \left( t - \frac{  \vert {\bf{r}} ^{\ast} - {\bf{r}} ' _{i} \vert  }{v}  \right)   \right]  , \label{AX} \\ A _{y}  ^{(1)} ({\bf{r}} , t ) &= - \frac{\tilde{\alpha}}{8 \pi } \frac{\mu}{\epsilon} \sqrt{\frac{\epsilon _{0}}{\mu _{0}}} \frac{I _{0} }{ \omega _{0} k _{0} (\vert z \vert + z _{0})}  \left( \frac{\partial ^{2}}{\partial x ^{2}} + k _{0} ^{2}  \right)   \int _{-d} ^{+d}  d x ^{\prime}  \sin  \left[  \omega _{0} \left( t - \frac{  \vert {\bf{r}} ^{\ast} - {\bf{r}} ' _{i} \vert }{v}  \right)   \right] , \label{AY}
\end{align}
 where ${\bf{r}} ' _{i} (x') = x' \hat{{\bf{e}}}_{x} - z _{0} \hat{{\bf{e}}}_{z} $ is the image point of  ${\bf{r}} ' (x')$ and ${\bf{r}} ^{\ast}  = x \hat{{\bf{e}}}_{x} + y \hat{{\bf{e}}}_{y} + \vert z \vert \hat{{\bf{e}}}_{z} $.   The first-order scalar potential in the time-domain is
 \begin{align}
\phi ^{(1)}  ({\bf{r}} ,  t )  =- I _{0} \frac{\tilde{\alpha}}{8 \pi } \frac{\mu}{\epsilon} \sqrt{\frac{\epsilon _{0}}{\mu _{0}}} \frac{y}{ \vert z \vert + z _{0} }    \int _{-d} ^{+d}  d x ^{\prime} \frac{ \sin  \left[ \omega _{0} \left( t - \frac{ \vert {\bf{r}} ^{\ast} - {\bf{r}} ' _{i} \vert  }{v}  \right)    \right] }{   \vert {\bf{r}} ^{\ast} - {\bf{r}} ' _{i} \vert  } .   \label{Scalar_Pot}
\end{align}
In the radiation zone these potentials can be further simplified. Indeed, one realizes than only the component $A_{y} ^{(1)}$ of the vector potential contributes to radiation (to be precise, the term proportional to $k_{0} ^{2}$ in Eq. (\ref{AY})), while the others decay faster than $1/r$.  Using the expansion $ \vert {\bf{r}} ^{\ast} - {\bf{r}} ' _{i} \vert   \approx    r ^{\ast  } - \frac{ x x' - z _{0} \vert z \vert  }{r ^{\ast  } } $ with $r ^{\ast  } = r$,  we get
\begin{align}
A _{y}  ^{(1)} ({\bf{r}} , t )  \approx   - \tilde{\alpha} \frac{I _{0}}{4 \pi \epsilon \omega _{0} }    \sqrt{\frac{\epsilon _{0}}{\mu _{0}}  \frac{\mu}{\epsilon} }  \frac{ r/x   }{  (\vert z \vert + z _{0})}    \frac{1}{v}   \sin  \left[ \omega _{0} \left( t -  \frac{r ^{2 } +  z _{0} \vert z \vert  }{v \, r  }   \right)    \right]  \sin    \left(   \frac{  \omega _{0}  d  }{v  } \frac{x}{r}  \right)   ,  \qquad  \phi ^{(1)}  ({\bf{r}} ,  t )  = v  \frac{y}{r} A _{y}  ^{(1)} ({\bf{r}} , t ) .
\end{align}
Keeping terms of the order $1/r$ the electromagnetic fields become
\begin{align}
{\bf{B}} ^{(1)}  =  - \tilde{\alpha} \frac{\mu I _{0}}{4  \pi }     \sqrt{\frac{\epsilon _{0}}{\mu _{0}}  \frac{\mu}{\epsilon} }  \cos  \left[ \omega _{0} \left( t -  \frac{r ^{2 } +  z _{0} \vert z \vert  }{v \, r  }   \right)    \right]  \sin    \left(   \frac{  \omega _{0}  d  }{v  } \frac{x}{r}  \right)   \frac{  r / x }{  \vert z \vert + z _{0} }     \hat{{\bf{e}}} _{y} \times \hat{{\bf{n}}}  , \qquad {\bf{E}} ^{(1)} = - v \, \hat{{\bf{n}}}  \times {\bf{B}} ^{(1)} .  \label{B1}
\end{align}

\begin{figure}
\includegraphics[scale=0.4]{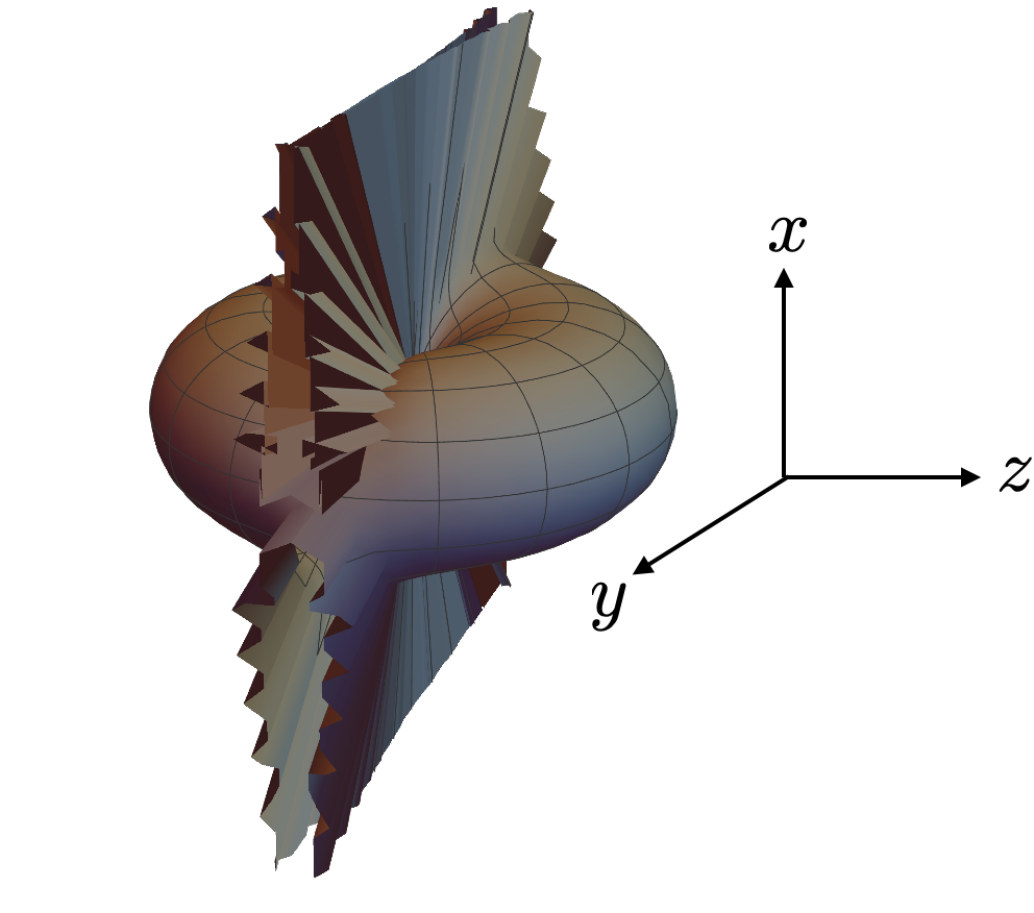}
\caption{Angular distribution of the radiated power by a thin linear antenna near a $\theta$ domain-wall. The position of the interface is evident.  } \label{Ang_Dist_Rad_Power}
\end{figure}

As shown in Ref. \cite{PhysRevD.92.125015}, energy conservation is unchanged in the presence of the $\theta$ term, i.e. given the Poynting vector ${\bf{S}}$ and energy density $u$ defined as usual ${\bf{S}} = {\bf{E}} \times {\bf{H}}$ and $u = \frac{1}{2} ( {\bf{E}} \cdot {\bf{D}} + {\bf{B}} \cdot {\bf{H}})$, respectively,  one realizes that $\partial _{t} u + \nabla \cdot {\bf{S}} = 0$. Momentum conservation is modified by a surface term due to the self-induced charge and current densities arising there, but it does not concern us in this paper. The angular distribution of the radiated power takes the usual form $\frac{dP}{d \Omega} =   \frac{v}{\mu}  r ^{2}   \vert {\bf{B}} \vert ^{2} $, with however a total magnetic field which acquires a contribution arising from the $\theta$-term. Keeping the lowest order in the magnetoelectric polarizability $\tilde{\alpha}$ and averaging (in time) over a complete cycle, the angular distribution of the radiated power by the linear antenna becomes
\begin{align}
\left< \frac{dP}{d \Omega } \right> \approx   \frac{v \mu I _{0} ^{2}}{8 \pi ^{2} }   \sin ^{2} \left( \frac{ \omega _{0} d    }{v}  \frac{x}{r} \right)   \left[   \frac{ 1 - (x/r) ^{2} }{ (x/r) ^{2} }   - 2 (\theta / \pi ) \sigma _{\mbox{\scriptsize Hall } }   \sqrt{  \frac{\mu}{\epsilon} }   \, \frac{y}{x} \frac{r}{ \vert z \vert   } \cos \left( \frac{z _{0} \omega}{v} \frac{z + \vert z \vert }{r} \right) \right] . \label{Radiated_Power}
\end{align}
The first term corresponds to the usual radiated power by a linear antenna embedded in an infinite material media described by the permittivity $\epsilon$ and permeability $\mu$.  The second term, which is proportional to the surface Hall conductivity, exhibits an interesting divergent behavior near the interface, as is depicted in Fig. \ref{Ang_Dist_Rad_Power}.  The physical origin of this term is understood from the surface Hall current and charge densities at the interface, which generates electromagnetic field which contribution to radiations in the near zone to the interface.

\section{Summary and discussion} \label{SummDisc}

Topological phases are an emerging class of materials which have attracted much attention in condensed matter physics. The remarkable microscopic properties encoded in the band structure of topological phases produces exceptional macroscopic electromagnetic response, which is described by a topological field theory akin to axion electrodynamics. In general, a linear magnetoelectric media can be described by adding the term $ \sim \theta _{ij} E_{i}B _{j}$ to the free enthalpy of the system, where $\theta _{ij}$ is the magnetoelectric tensor. In this frame, axion electrodynamics is a particular case when the ME tensor is of the form $\theta _{ij} = \theta \delta _{ij}$.

In this paper we derived, in a perturbative fashion, solutions to the time-dependent axion electrodynamics field equations for a picewise constant axion angle. The validity of the perturbative program relies in the fact that the axion coupling, in the case of topological insulators, is proportional to the fine structure constant. In the particle physics context this condition is vastly fulfilled, since it is expected to be very small. The difference is that in the former case the presence of boundaries is a natural task, while in the latter the axion is assumed to  be distributed over the whole space.

To validate our perturbative scheme, we first tackle the static axion field equations and derive the components of the Green's matrix up to second order in the axion coupling. We show that our results are consistent with the previously reported exact Green's matrix when Taylor expanded up to second order in the coupling. We discuss in passing the image magnetic monopole effect, which is a salient manifestation of the unconventional magnetoelectric response of topological insulators. In the time-dependent case, the axion field equations become more involved and no analytical results have been reported yet, although there are some first attempts towards this direction. Here we aim to fill in this gap by applying the same perturbative program to the axion field equations. For the sake of simplicity, in this paper we choose that the bulks have the same optical properties, namely, $\epsilon _{1} = \epsilon _{2} \equiv \epsilon$ and $\mu _{1} = \mu _{2} \equiv \mu$, which is made to highlight and isolate the effects of the axion coupling. In a forthcoming paper we will relax this condition and generalize our results to the case of two topologically insulating media characterized by different nontrivial optical parameters. This generalization is of course far from trivial. To illustrate the importance of the axion angle in the time-dependent case considered here, we investigate the radiation of a short linear antenna above a topologically insulating domain-wall with different values of $\theta$. 

We first computed numerically the electromagnetic fields due to the linear antenna by using the exact Green's function and verify that they exhibit the expected behavior. We interpret the induced fields in terms of the Hall current at the interface between the two half-spaces characterized by different $\theta$-angles. Next, to investigate the effects of the $\theta$-term in the power radiated, we use the asymptotic expression for the Green's function and evaluate the fields in the radiation zone. These field allow us to compute the time-averaged angular distribution of the radiated power, which as expected, is dominated by the usual Maxwellian term, but becomes corrected due to the $\theta$-term. Such correction is quite interesting, since exhibits a divergent-type behavior near the interface, which arises from the field produced by the surface charge and current densities there.  This is a distinctive behavior with respect to the usual radiation pattern. 

In summary, a general solution to the static axion field equations has been previously elaborated in terms of Green's function techniques; however, the time-dependent case has received less attention mainly due to its complexity. In this paper we introduced a perturbative scheme which was successfully validated with the static theory and then we extended the analysis to the time-dependent case. We obtained general expressions for the corresponding matrix of Green's function which we also approximated in the far-field regime. We expect our method and results will be of considerable relevance and that they may constitute the basis for numerous other researches which include more intricate model of antennas and the radiation of particles near the $\theta$-interface.

\acknowledgements{M. I.-M. was supported by the CONAHCyT fellowship No. 843426.  A. M.-R. and M. I.-M. has been partially supported by DGAPA-UNAM Project No. IA102722 and by Project CONAHCyT (M\'{e}xico) No. 428214. Useful discussions with L. F. Urrutia, M. Cambiaso and L. Medel are warmly appreciated.}

\section*{Data Availability Statement \; } No datasets were generated or analyzed during the current study.

\section*{Declarations \; }

\textbf{Conflict of interest} \; The authors declare no conflict of interest.

\appendix

\section{Static Green's function} \label{Detailed-calculations}

\subsection{First-order Green's functions}

Let us consider the first-order corrections. The solution to Eq. (\ref{n-Order-Static}) are
\begin{align}
    \phi ^{(1)} ({\bf{r}}) = \int G ^{(1)} _{0i} ({\bf{r}},{\bf{r}} ^{\prime}) \, J _{i} ({\bf{r}} ^{\prime}) \, d ^{3} {\bf{r}} ^{\prime} , \qquad A ^{(1)} _{i} ({\bf{r}}) = \int G ^{(1)} _{i0} ({\bf{r}},{\bf{r}} ^{\prime}) \, \rho ({\bf{r}} ^{\prime}) \, d ^{3} {\bf{r}} ^{\prime} , 
\end{align}
where
\begin{align}
    G ^{(1)} _{0i} ({\bf{r}},{\bf{r}} ^{\prime}) &= - \tilde{\alpha} \, \sqrt{\frac{\epsilon _{0}}{\mu _{0}}} \, \int G _{\epsilon} ({\bf{r}},{\bf{r}} ^{\prime \prime}) \, \delta (z ^{\prime \prime}) \, \epsilon _{zji} \, \partial _{j} ^{\prime \prime} \, G _{\mu} ^{( \perp )} ({\bf{r}} ^{\prime \prime},{\bf{r}} \,  ^{\prime}) \, d ^{3} {\bf{r}} ^{\prime \prime} ,  \\ G ^{(1)} _{i0} ({\bf{r}},{\bf{r}} ^{\prime}) &= - \tilde{\alpha} \, \sqrt{\frac{\epsilon _{0}}{\mu _{0}}} \, \int G _{\mu} ^{( \perp )} ({\bf{r}},{\bf{r}} ^{\prime \prime}) \, \delta (z ^{\prime \prime}) \, \epsilon _{zji} \, \partial _{j} ^{\prime \prime} \, G _{\epsilon}  ({\bf{r}} ^{\prime \prime},{\bf{r}} \,  ^{\prime}) \, d ^{3} {\bf{r}} ^{\prime \prime} .
\end{align}
Noting that partial derivatives apply only in the transverse coordinates $x ^{\prime \prime}$ and $y ^{\prime \prime}$, the properties of the Dirac delta allow us to compute the integrals with respect to the $z ^{\prime \prime}$-coordinate in a simple fashion. Now, if we use $\partial _{j} ^{\prime \prime} \, G _{\mu} ^{( \perp )} ({\bf{r}} ^{\prime \prime},{\bf{r}} \,  ^{\prime}) = - \partial _{j} ^{\prime} \, G _{\mu} ^{( \perp )} ({\bf{r}} ^{\prime \prime},{\bf{r}} \,  ^{\prime})$ and $\partial _{j} ^{\prime \prime} \, G _{\epsilon}  ({\bf{r}} ^{\prime \prime},{\bf{r}} \,  ^{\prime}) = - \partial _{j} ^{\prime} \, G _{\epsilon}  ({\bf{r}} ^{\prime \prime},{\bf{r}} \,  ^{\prime})$ for $j=x,y$, the double primed derivatives can be get out of the integrals as a minus primed derivative. So we are left with
\begin{align}
    G ^{(1)} _{0i} ({\bf{r}},{\bf{r}} ^{\prime}) &= \tilde{\alpha} \, \sqrt{\frac{\epsilon _{0}}{\mu _{0}}} \, \epsilon _{zji} \, \partial _{j} ^{\prime} \, \int _{\Sigma} G _{\mu} ^{( \perp )} ({\bf{r}},{\bf{r}} ^{\prime \prime} _{\perp} ) \,  G _{\epsilon}  ({\bf{r}} ^{\prime \prime} _{\perp},{\bf{r}} \,  ^{\prime}) \, d ^{2} {\bf{r}} ^{\prime \prime} _{\perp} , \label{IntG0i(0)} \\ G ^{(1)} _{i0} ({\bf{r}},{\bf{r}} ^{\prime}) &= \tilde{\alpha} \, \sqrt{\frac{\epsilon _{0}}{\mu _{0}}} \, \epsilon _{zji} \, \partial _{j} ^{\prime} \, \int _{\Sigma} G _{\epsilon} ({\bf{r}},{\bf{r}} ^{\prime \prime} _{\perp} ) \,  G _{\mu} ^{( \perp )}   ({\bf{r}} ^{\prime \prime} _{\perp},{\bf{r}} \,  ^{\prime}) \, d ^{2} {\bf{r}} ^{\prime \prime} _{\perp} . \label{IntGi0(0)}
\end{align} 
Using the electric (\ref{Electric-GF}) and magnetic (\ref{xi-Magnetic-GF}) Green functions we realize that
\begin{align}
    G ^{(1)} _{0i} ({\bf{r}},{\bf{r}} ^{\prime}) = G ^{(1)} _{i0} ({\bf{r}},{\bf{r}} ^{\prime}) = \frac{\tilde{\alpha} \sqrt{\epsilon _{0} / \mu _{0}} }{( \epsilon _{1} + \epsilon _{2} )( 1/ \mu _{1} + 1 / \mu _{2} )}  \, \epsilon _{zji} \, \partial _{j} ^{\prime} \, \int _{\Sigma} \frac{1}{ \sqrt{ \vert {\bf{r}} _{\perp } - {\bf{r}} ^{\prime \prime} {} _{\!\! \perp } \vert ^{2} + z ^{2}} } \,  \frac{1}{ \sqrt{ \vert {\bf{r}} _{\perp } ^{\prime \prime} - {\bf{r}} ^{\prime} {} _{\!\! \perp } \vert ^{2} + z ^{\prime \, 2}} } \, \frac{d ^{2} {\bf{r}} ^{\prime \prime} _{\perp}}{(2 \pi ) ^{2}} .  \label{IntG0i(1)}
\end{align}
To carry out integration, we use the following integral representation:
\begin{align}
\frac{1}{\vert {\bf{r}} - {\bf{r}} ^{\prime} \vert } = 4 \pi \int _{\Sigma _{{\bf{k}} _{\perp}} } \frac{d ^{2} {\bf{k}} _{\perp} }{(2 \pi ) ^{2}} \; e ^{i {\bf{k}} _{\perp} \cdot ({\bf{r}} _{\perp} - {\bf{r}} _{\perp} ^{\prime} )} \,  \frac{e ^{- k _{\perp} \vert z - z ^{\prime} \vert }  }{2 k _{\perp}} ,  \label{IntegralRep}
\end{align}
where $\Sigma _{{\bf{k}} _{\perp}}$ is the two-dimensional space $k_{x}$-$k_{y}$. Using this representation twice in Eq. (\ref{IntG0i(1)}) we get
\begin{align}
    G ^{(1)} _{0i} ({\bf{r}},{\bf{r}} ^{\prime}) &= G ^{(1)} _{i0} ({\bf{r}},{\bf{r}} ^{\prime}) = \frac{\tilde{\alpha} \sqrt{\epsilon _{0} / \mu _{0}} }{( \epsilon _{1} + \epsilon _{2} )( 1/ \mu _{1} + 1 / \mu _{2} )}  \, \epsilon _{zji} \notag \\ & \hspace{3cm} \times \partial _{j} ^{\prime} \, \int _{\Sigma _{{\bf{k}} _{\perp}} } \frac{d ^{2} {\bf{k}} _{\perp} }{2 \pi } \;  \frac{e ^{- k _{\perp} \vert z  \vert }  }{ k _{\perp}} \, \int _{\Sigma _{{\bf{k}} _{\perp} ^{\prime}} } \frac{d ^{2} {\bf{k}} _{\perp} ^{\prime} }{ 2 \pi } \;   \frac{e ^{- k _{\perp} ^{\prime} z ^{\prime} }  }{k _{\perp} ^{\prime}} \,  \int _{\Sigma}  \, e ^{i {\bf{k}} _{\perp} \cdot ({\bf{r}} _{\perp} - {\bf{r}} _{\perp} ^{\prime \prime} )} \, e ^{i {\bf{k}} _{\perp} ^{\prime} \cdot ({\bf{r}} _{\perp} ^{\prime \prime} - {\bf{r}} _{\perp} ^{\prime } )}  \, \frac{d ^{2} {\bf{r}} ^{\prime \prime} _{\perp}}{(2 \pi ) ^{2}} .  \label{IntG0i(1)-2}
\end{align}
Integration with respect to the coordinates ${\bf{r}} ^{\prime \prime} _{\perp}$ produces a Dirac delta on the momenta ${\bf{k}} _{\perp}$ and ${\bf{k}} _{\perp} ^{\prime}$, i.e.
\begin{align}
    G ^{(1)} _{0i} ({\bf{r}},{\bf{r}} ^{\prime}) = G ^{(1)} _{i0} ({\bf{r}},{\bf{r}} ^{\prime}) &= \frac{\tilde{\alpha} \sqrt{\epsilon _{0} / \mu _{0}} }{( \epsilon _{1} + \epsilon _{2} )( 1/ \mu _{1} + 1 / \mu _{2} )}  \, \epsilon _{zji}  \notag \\ & \hspace{2cm} \times \partial _{j} ^{\prime} \, \int _{\Sigma _{{\bf{k}} _{\perp}} } \frac{d ^{2} {\bf{k}} _{\perp}}{2 \pi } \;  \frac{e ^{- k _{\perp} \vert z  \vert }  }{ k _{\perp}} \, \int _{\Sigma _{{\bf{k}} _{\perp} ^{\prime}} } \frac{d ^{2} {\bf{k}} _{\perp} ^{\prime}}{2 \pi } \;   \frac{e ^{- k _{\perp} ^{\prime} z ^{\prime} }  }{k _{\perp} ^{\prime}} \,   e ^{i {\bf{k}} _{\perp} \cdot {\bf{r}} _{\perp} - i {\bf{k}} _{\perp} ^{\prime} \cdot {\bf{r}} _{\perp} ^{\prime } } \,  \delta ({\bf{k}} _{\perp} ^{\prime} - {\bf{k}} _{\perp} ) \notag \\ &=  \frac{\tilde{\alpha} \sqrt{\epsilon _{0} / \mu _{0}} }{( \epsilon _{1} + \epsilon _{2} )( 1/ \mu _{1} + 1 / \mu _{2} )}  \, \epsilon _{zji}  \; \partial _{j} ^{\prime} \, \int _{\Sigma _{{\bf{k}} _{\perp}} } \frac{d ^{2} {\bf{k}} _{\perp}}{(2 \pi ) ^{2}} \; e ^{i {\bf{k}} _{\perp} \cdot ( {\bf{r}} _{\perp} - {\bf{r}} _{\perp} ^{\prime } ) } \; \frac{e ^{- k _{\perp} ( \vert z  \vert + z ^{\prime} ) }  }{ k _{\perp} ^{2}}  ,  \label{IntG0i(1)-3}
\end{align}
where we used the properties of the Dirac delta to perform integration over ${\bf{k}} _{\perp} ^{\prime}$. The resulting double integral becomes easier to perform if we express the area element in polar coordinates, i.e. $d ^{2} {\bf{k}} _{\perp} = k _{\perp} dk _{\perp} d \varphi$, and choose the $k_{x}$ axis in the direction of the vector ${\bf{r}} _{\perp} - {\bf{r}} _{\perp} ^{\prime }$. Noting that ${\bf{k}} _{\perp} \cdot ( {\bf{r}} _{\perp} - {\bf{r}} _{\perp} ^{\prime } ) = \vert {\bf{r}} _{\perp} - {\bf{r}} _{\perp} ^{\prime } \vert \, k _{\perp} \cos \varphi $, we can write 
\begin{align}
    G ^{(1)} _{0i} ({\bf{r}},{\bf{r}} ^{\prime}) = G ^{(1)} _{i0} ({\bf{r}},{\bf{r}} ^{\prime}) &=  \frac{\tilde{\alpha} \sqrt{\epsilon _{0} / \mu _{0}} }{2 \pi ( \epsilon _{1} + \epsilon _{2} )( 1/ \mu _{1} + 1 / \mu _{2} )}  \, \epsilon _{zji}  \; \partial _{j} ^{\prime} \, \int _{0} ^{\infty}  dk _{\perp}  \, \frac{e ^{- k _{\perp} ( \vert z  \vert + z ^{\prime}  ) }  }{ k _{\perp} } \left( \frac{1}{2 \pi } \int _{0} ^{2 \pi } d \varphi \, e ^{i \vert {\bf{r}} _{\perp} - {\bf{r}} _{\perp} ^{\prime } \vert \, k _{\perp} \cos \varphi }   \right)  .   \label{IntG0i(1)-4}
\end{align}
The parentheses in this equation enclose an integral representation of the Bessel function $J _{0}$. The resulting integral is
\begin{align}
    G ^{(1)} _{0i} ({\bf{r}},{\bf{r}} ^{\prime}) = G ^{(1)} _{i0} ({\bf{r}},{\bf{r}} ^{\prime}) &= \frac{\tilde{\alpha} \sqrt{\epsilon _{0} / \mu _{0}} }{2 \pi ( \epsilon _{1} + \epsilon _{2} )( 1/ \mu _{1} + 1 / \mu _{2} )}  \, \epsilon _{zji}  \; \partial _{j} ^{\prime} \, \int _{0} ^{\infty}  dk _{\perp}  \, J _{0} ( \vert {\bf{r}} _{\perp} - {\bf{r}} _{\perp} ^{\prime } \vert \, k _{\perp} ) \, \frac{e ^{- k _{\perp} ( \vert z  \vert + z ^{\prime}  ) }  }{ k _{\perp} } ,   \label{IntG0i(1)-5}
\end{align}
and taking the primed derivative
\begin{align}
    G ^{(1)} _{0i} ({\bf{r}},{\bf{r}} ^{\prime}) = G ^{(1)} _{i0} ({\bf{r}},{\bf{r}} ^{\prime}) &= \frac{\tilde{\alpha} \sqrt{\epsilon _{0} / \mu _{0}} }{2 \pi ( \epsilon _{1} + \epsilon _{2} )( 1/ \mu _{1} + 1 / \mu _{2} )}  \, \frac{ \epsilon _{zji}  \, ( {\bf{r}} _{\perp} - {\bf{r}} _{\perp} ^{\prime } ) _{j} }{\vert {\bf{r}} _{\perp} - {\bf{r}} _{\perp} ^{\prime } \vert } \, \int _{0} ^{\infty}  dk _{\perp}  \, J _{1} ( \vert {\bf{r}} _{\perp} - {\bf{r}} _{\perp} ^{\prime } \vert \, k _{\perp} ) \, e ^{- k _{\perp} ( \vert z  \vert + z ^{\prime} ) } .  \label{IntG0i(1)-6}
\end{align}
This integral is well known \cite{gradshteyn2007}. The final result is then
\begin{align}
    G ^{(1)} _{0i} ({\bf{r}},{\bf{r}} ^{\prime}) = G ^{(1)} _{i0} ({\bf{r}},{\bf{r}} ^{\prime}) &= \frac{\tilde{\alpha} \sqrt{\epsilon _{0} / \mu _{0}} }{2 \pi ( \epsilon _{1} + \epsilon _{2} )( 1/ \mu _{1} + 1 / \mu _{2} )}  \, \frac{ \epsilon _{zji}  \, ( {\bf{r}} _{\perp} - {\bf{r}} _{\perp} ^{\prime } ) _{j} }{\vert {\bf{r}} _{\perp} - {\bf{r}} _{\perp} ^{\prime } \vert ^{2}} \, \left[ 1 - \frac{\vert z  \vert + z ^{\prime} }{ \sqrt{ \vert {\bf{r}} _{\perp} - {\bf{r}} _{\perp} ^{\prime } \vert ^{2} + ( \vert z  \vert + z ^{\prime} ) ^{2} }} \right] .  \label{IntG0i(1)-Fin}
\end{align}

\subsection{Second-order Green's functions}

We now turn to the second-order corrections to the GFs. Integration of Eq. (\ref{n-Order-Static}) produces
\begin{align}
    \phi ^{(2)} ({\bf{r}}) = \int G ^{(2)} _{00} ({\bf{r}},{\bf{r}} ^{\prime}) \, \rho ({\bf{r}} ^{\prime}) \, d ^{3} {\bf{r}} ^{\prime} , \qquad A ^{(2)} _{i} ({\bf{r}}) = \int G ^{(2)} _{ij} ({\bf{r}},{\bf{r}} ^{\prime}) \, J_{j} ({\bf{r}} ^{\prime}) \, d ^{3} {\bf{r}} ^{\prime} , 
\end{align}
where
\begin{align}
    G ^{(2)} _{00} ({\bf{r}},{\bf{r}} ^{\prime}) &= - \tilde{\alpha} \, \sqrt{\frac{\epsilon _{0}}{\mu _{0}}} \, \int G _{\epsilon} ({\bf{r}},{\bf{r}} ^{\prime \prime}) \, \delta (z ^{\prime \prime}) \, \epsilon _{zji} \, \partial _{j} ^{\prime \prime} \, G _{i0} ^{( 1 )} ({\bf{r}} ^{\prime \prime},{\bf{r}} \,  ^{\prime}) \, d ^{3} {\bf{r}} ^{\prime \prime} ,  \\ G ^{(2)} _{ij} ({\bf{r}},{\bf{r}} ^{\prime}) &= - \tilde{\alpha} \, \sqrt{\frac{\epsilon _{0}}{\mu _{0}}} \, \int G _{\mu} ^{( \perp )} ({\bf{r}},{\bf{r}} ^{\prime \prime}) \, \delta (z ^{\prime \prime}) \, \epsilon _{zki} \, \partial _{k} ^{\prime \prime} \, G _{0j} ^{(1)}  ({\bf{r}} ^{\prime \prime},{\bf{r}} \,  ^{\prime}) \, d ^{3} {\bf{r}} ^{\prime \prime} . 
\end{align}
Using the first-order Green's function given by Eq. (\ref{IntG0i(1)-Fin}), we verify that $\partial _{j} ^{\prime \prime} \, G _{i0} ^{( 1 )} ({\bf{r}} ^{\prime \prime},{\bf{r}} \,  ^{\prime}) = - \partial _{j} ^{\prime} \, G _{i0} ^{( 1 )} ({\bf{r}} ^{\prime \prime},{\bf{r}} \,  ^{\prime})$ for $j=1,2$, and hence the double primed derivatives can be get out of the integrals as a minus primed derivative, as before. Performing the integration with respect to $z ^{\prime \prime}$ we get
\begin{align}
    G ^{(2)} _{00} ({\bf{r}},{\bf{r}} ^{\prime}) &= \tilde{\alpha} \, \sqrt{\frac{\epsilon _{0}}{\mu _{0}}} \, \epsilon _{zji} \, \partial _{j} ^{\prime} \, \int _{\Sigma} G _{\epsilon} ({\bf{r}},{\bf{r}} ^{\prime \prime} _{\perp} ) \, G _{i0} ^{( 1 )} ({\bf{r}} ^{\prime \prime} _{\perp} ,{\bf{r}} \,  ^{\prime}) \, d ^{2} {\bf{r}} ^{\prime \prime} _{\perp} ,  \\ G ^{(2)} _{ij} ({\bf{r}},{\bf{r}} ^{\prime}) &= \tilde{\alpha} \, \sqrt{\frac{\epsilon _{0}}{\mu _{0}}} \, \, \epsilon _{zki} \, \partial _{k} ^{\prime}  \int _{\Sigma} G _{\mu} ^{( \perp )} ({\bf{r}},{\bf{r}} ^{\prime \prime} _{\perp} ) \, G _{0j} ^{(1)}  ({\bf{r}} ^{\prime \prime} _{\perp} ,{\bf{r}} \,  ^{\prime}) \, d ^{2} {\bf{r}} ^{\prime \prime} _{\perp} . 
\end{align}
Using the integral representations (\ref{IntG0i(0)}) for the $0i$-components and (\ref{IntGi0(0)}) we find
\begin{align}
    G ^{(2)} _{00} ({\bf{r}},{\bf{r}} ^{\prime}) &= ( \tilde{\alpha} \sqrt{\epsilon _{0} / \mu _{0}} \, ) ^{2} \, \epsilon _{zji}  \, \epsilon _{zki} \, \partial _{j} ^{\prime} \,  \partial _{k} ^{\prime} \, \int _{\Sigma} \int _{\Sigma} G _{\epsilon} ({\bf{r}},{\bf{r}} ^{\prime \prime} _{\perp} ) \,  G _{\epsilon} ({\bf{r}} ^{\prime \prime} _{\perp} ,{\bf{r}} ^{\prime \prime \prime} _{\perp} ) \,  G _{\mu} ^{( \perp )}   ({\bf{r}} ^{\prime \prime \prime } _{\perp},{\bf{r}} \,  ^{\prime}) \, d ^{2} {\bf{r}} ^{\prime \prime \prime} _{\perp}  \, d ^{2} {\bf{r}} ^{\prime \prime} _{\perp} ,  \\ G ^{(2)} _{ij} ({\bf{r}},{\bf{r}} ^{\prime}) &= ( \tilde{\alpha} \sqrt{\epsilon _{0} / \mu _{0}} \, ) ^{2} \, \epsilon _{zki} \,  \epsilon _{zlj} \, \partial _{l} ^{\prime} \, \partial _{k} ^{\prime} \, \int _{\Sigma} \int _{\Sigma} G _{\mu} ^{( \perp )} ({\bf{r}},{\bf{r}} ^{\prime \prime} _{\perp} ) \,  G _{\epsilon} ({\bf{r}} ^{\prime \prime} _{\perp} ,{\bf{r}} ^{\prime \prime \prime} _{\perp} ) \,  G _{\mu} ^{( \perp )}   ({\bf{r}} ^{\prime \prime \prime } _{\perp},{\bf{r}} \,  ^{\prime}) \, d ^{2} {\bf{r}} ^{\prime \prime \prime} _{\perp} \, d ^{2} {\bf{r}} ^{\prime \prime} _{\perp} ,
\end{align}
and substituting the electric (\ref{Electric-GF}) and magnetic (\ref{xi-Magnetic-GF}) Green functions we get
\begin{align}
    G ^{(2)} _{00} ({\bf{r}},{\bf{r}} ^{\prime}) &= 
    \frac{( \tilde{\alpha} \sqrt{\epsilon _{0} / \mu _{0}} \, ) ^{2}}{2 \pi ( \epsilon _{1} + \epsilon _{2} ) ^{2} (1 / \mu _{1} + 1 / \mu _{2}) } \; \mbox{tr} \, [ {\bf{I}} ({\bf{r}},{\bf{r}} ^{\prime}) ] ,  \\ G ^{(2)} _{ij} ({\bf{r}},{\bf{r}} ^{\prime}) &= \frac{( \tilde{\alpha} \sqrt{\epsilon _{0} / \mu _{0}} \, ) ^{2}}{2 \pi ( \epsilon _{1} + \epsilon _{2} ) (1 / \mu _{1} + 1 / \mu _{2}) ^{2} }   \; I _{ij} ({\bf{r}},{\bf{r}} ^{\prime}) ,
\end{align}
where we have defined the symmetric matrix ${\bf{I}} = (I _{ij}) \in \mathbb{R} ^{2 \times 2} $, with components
\begin{align}
    I _{ij} ({\bf{r}},{\bf{r}} ^{\prime}) = (\delta _{ij} \delta _{lm} - \delta _{il} \delta _{jm} ) \, \partial _{l} ^{\prime} \, \partial _{m} ^{\prime} \, \int _{\Sigma} \int _{\Sigma} \, \frac{1}{ \sqrt{ \vert {\bf{r}} _{\perp } - {\bf{r}} ^{\prime \prime} {} _{\!\! \perp } \vert ^{2} + z ^{2}} } \frac{1}{ \sqrt{ \vert {\bf{r}} _{\perp } ^{\prime \prime} - {\bf{r}} ^{\prime \prime \prime} {} _{\!\! \perp } \vert ^{2} + 0 ^{2}} }   \frac{1}{ \sqrt{ \vert {\bf{r}} _{\perp } ^{\prime \prime \prime} - {\bf{r}} ^{\prime} {} _{\!\! \perp } \vert ^{2} + z ^{ \prime \, 2}} } \, \frac{d ^{2} {\bf{r}} ^{\prime \prime \prime} _{\perp}}{2 \pi} \, \frac{d ^{2} {\bf{r}} ^{\prime \prime} _{\perp}}{2 \pi}
\end{align}
To evaluate this expression, we use thrice the integral representation (\ref{IntegralRep}), i.e. 
\begin{align}
I _{ij} ({\bf{r}},{\bf{r}} ^{\prime}) &= (\delta _{ij} \delta _{lm} - \delta _{il} \delta _{jm} ) \, \partial _{l} ^{\prime} \, \partial _{m} ^{\prime} \, \int _{\Sigma} \int _{\Sigma} \, \left( \int _{\Sigma _{{\bf{k}} _{\perp}} } \frac{d^{2} {\bf{k}} _{\perp}}{2 \pi} e ^{i {\bf{k}} _{\perp} \cdot ({\bf{r}} _{\perp } - {\bf{r}} ^{\prime \prime} _{\perp} ) } \frac{e ^{- k _{\perp} \vert z \vert }}{k _{\perp}} \right)  \, \left( \int _{\Sigma _{{\bf{k}} _{\perp}} } \frac{d^{2} {\bf{k}} _{\perp} ^{\prime}}{2 \pi} e ^{i {\bf{k}} _{\perp} ^{\prime} \cdot ({\bf{r}} _{\perp } ^{\prime \prime} - {\bf{r}} ^{\prime \prime \prime} _{\perp} ) } \frac{e ^{- k _{\perp} ^{\prime} (0) }}{k _{\perp} ^{\prime}} \right) \notag \\ & \hspace{7cm} \left( \int _{\Sigma _{{\bf{k}} _{\perp}} } \frac{d^{2} {\bf{k}} _{\perp} ^{\prime \prime} }{2 \pi} e ^{i {\bf{k}} _{\perp} ^{\prime \prime} \cdot ({\bf{r}} _{\perp } ^{\prime \prime \prime} - {\bf{r}} ^{\prime} _{\perp} ) } \frac{e ^{- k _{\perp} ^{\prime \prime} z ^{\prime} }}{k _{\perp} ^{\prime \prime} } \right) \, \frac{d ^{2} {\bf{r}} ^{\prime \prime \prime} _{\perp}}{2 \pi} \, \frac{d ^{2} {\bf{r}} ^{\prime \prime} _{\perp}}{2 \pi} .
\end{align}
Integrating with respect to ${\bf{r}} ^{\prime \prime} _{\perp}$ and ${\bf{r}} ^{\prime \prime \prime} _{\perp}$ produces two Dirac deltas in momentum:
\begin{align}
    I _{ij} ({\bf{r}},{\bf{r}} ^{\prime}) &= (\delta _{ij} \delta _{lm} - \delta _{il} \delta _{jm} ) \, \partial _{l} ^{\prime} \, \partial _{m} ^{\prime} \, \int _{\Sigma _{{\bf{k}} _{\perp}} } \int _{\Sigma _{{\bf{k}} _{\perp}} } \int _{\Sigma _{{\bf{k}} _{\perp}} } \frac{d ^{2} {\bf{k}} _{\perp} \, d^{2} {\bf{k}} _{\perp} ^{\prime} \, d^{2} {\bf{k}} _{\perp} ^{\prime \prime} }{2 \pi}  \frac{ e ^{- k _{\perp} \vert z \vert } e ^{- k _{\perp} ^{\prime \prime} z ^{\prime} }}{k _{\perp} k _{\perp} ^{\prime} k _{\perp} ^{\prime \prime}  }  e ^{ i {\bf{k}} _{\perp} \cdot {\bf{r}} _{\perp } - i {\bf{k}} _{\perp} ^{\prime \prime } \cdot {\bf{r}} _{\perp } ^{\prime} } \, \delta ({\bf{k}} _{\perp} ^{\prime} - {\bf{k}} _{\perp} ) \, \delta ({\bf{k}} _{\perp} ^{\prime \prime} - {\bf{k}} _{\perp} ^{\prime }) \notag \\ &= (\delta _{ij} \delta _{lm} - \delta _{il} \delta _{jm} ) \, \partial _{l} ^{\prime} \, \partial _{m} ^{\prime} \, \int _{\Sigma _{{\bf{k}} _{\perp}} } \frac{ d ^{2} {\bf{k}} _{\perp} }{2 \pi} \frac{e ^{- k _{\perp} ( \vert z \vert  + z ^{\prime} ) }}{k _{\perp} ^{3}} e ^{ i {\bf{k}} _{\perp} \cdot ( {\bf{r}} _{\perp }  - {\bf{r}} _{\perp } ^{\prime} ) } 
\end{align}
where we have used the properties of the Dirac delta. We now introduce the polar coordinates $(k _{x}, k _{y} ) = k _{\perp} ( \cos \varphi , \sin \varphi ) $ and choose the vector ${\bf{r}} _{\perp }  - {\bf{r}} _{\perp } ^{\prime}$ along the $k _{x}$ direction. So, the angular integral yields
\begin{align}
    I _{ij} ({\bf{r}},{\bf{r}} ^{\prime}) &= (\delta _{ij} \delta _{lm} - \delta _{il} \delta _{jm} ) \, \partial _{l} ^{\prime} \, \partial _{m} ^{\prime} \, \int _{0} ^{\infty} d  k _{\perp} \,  \frac{e ^{- k _{\perp} ( \vert z \vert  + z ^{\prime} ) }}{k _{\perp} ^{2}} J _{0} ( \, \vert {\bf{r}} _{\perp }  - {\bf{r}} _{\perp } ^{\prime} \vert k _{\perp}  \, ) . 
\end{align}
This integral does not converge as is, but its derivatives does. Taking one of the derivatives we find
\begin{align}
    I _{ij} ({\bf{r}},{\bf{r}} ^{\prime}) &= (\delta _{ij} \delta _{lm} - \delta _{il} \delta _{jm} ) \, \partial _{l} ^{\prime} \, \left[  \frac{({\bf{r}} _{\perp }  - {\bf{r}} _{\perp } ^{\prime}) _{m} }{\vert {\bf{r}} _{\perp }  - {\bf{r}} _{\perp } ^{\prime} \vert \; } \, \int _{0} ^{\infty} d  k _{\perp} \,  \frac{e ^{- k _{\perp} ( \vert z \vert  + z ^{\prime} ) }}{k _{\perp}} J _{1} ( \, \vert {\bf{r}} _{\perp }  - {\bf{r}} _{\perp } ^{\prime} \vert k _{\perp}  \, ) \right] . 
\end{align}
To perform this integral in a convergent manner, we use the following trick: $\int _{\Lambda} ^{\infty} e ^{-k _{\perp} \lambda} d \lambda = e ^{- k _{\perp} \Lambda} / k _{\perp}$, and therefore
\begin{align}
    I _{ij} ({\bf{r}},{\bf{r}} ^{\prime}) &= (\delta _{ij} \delta _{lm} - \delta _{il} \delta _{jm} ) \, \partial _{l} ^{\prime} \, \left[  \frac{({\bf{r}} _{\perp }  - {\bf{r}} _{\perp } ^{\prime}) _{m} }{\vert {\bf{r}} _{\perp }  - {\bf{r}} _{\perp } ^{\prime} \vert \; } \, \int _{\vert z \vert  + z ^{\prime}} ^{\infty} d \lambda \int _{0} ^{\infty} d  k _{\perp} \,  e ^{- k _{\perp} \lambda } \, J _{1} ( \, \vert {\bf{r}} _{\perp }  - {\bf{r}} _{\perp } ^{\prime} \vert k _{\perp}  \, ) \right] ,
\end{align}
and this integral is well-known; see, for example, Ref. \cite{gradshteyn2007}. The result is
\begin{align}
    I _{ij} ({\bf{r}},{\bf{r}} ^{\prime}) &= (\delta _{ij} \delta _{lm} - \delta _{il} \delta _{jm} ) \, \partial _{l} ^{\prime} \, \left[  \frac{({\bf{r}} _{\perp }  - {\bf{r}} _{\perp } ^{\prime}) _{m}}{\vert {\bf{r}} _{\perp }  - {\bf{r}} _{\perp } ^{\prime} \vert ^{2} } \, \int _{\vert z \vert  + z ^{\prime}} ^{\infty} d \lambda \left( 1 - \frac{\lambda}{\sqrt{\vert {\bf{r}} _{\perp }  - {\bf{r}} _{\perp } ^{\prime} \vert ^{2} + \lambda ^{2} }} \right) \right] ,
\end{align}
from which we finally obtain
\begin{align}
    I _{ij} ({\bf{r}},{\bf{r}} ^{\prime}) &= (\delta _{ij} \delta _{lm} - \delta _{il} \delta _{jm} ) \, \partial _{l} ^{\prime} \, \left[  \frac{({\bf{r}} _{\perp }  - {\bf{r}} _{\perp } ^{\prime}) _{m}}{\vert {\bf{r}} _{\perp }  - {\bf{r}} _{\perp } ^{\prime} \vert ^{2} } \, \left( \sqrt{\vert {\bf{r}} _{\perp }  - {\bf{r}} _{\perp } ^{\prime} \vert ^{2} + (\vert z \vert  + z ^{\prime}) ^{2} } - (\vert z \vert  + z ^{\prime}) \right) \right] .
\end{align}
From this expression we find
\begin{align}
    \mbox{tr} \, [ {\bf{I}} ({\bf{r}},{\bf{r}} ^{\prime}) ] =  \frac{1}{ \sqrt{ \vert {\bf{r}} _{\perp } - {\bf{r}} ^{\prime} {} _{\!\! \perp } \vert ^{2} + (\vert z \vert  + z ^{\prime}) ^{2} } } ,
\end{align}
and
\begin{align}
    I _{ij} ({\bf{r}},{\bf{r}} ^{\prime}) &=  \delta _{ij} \, \mbox{tr} \, [ {\bf{I}} ({\bf{r}},{\bf{r}} ^{\prime}) ] + \partial _{i} \, K _{j} ({\bf{r}},{\bf{r}} ^{\prime}) , 
\end{align}
where
\begin{align}
    K _{j} ({\bf{r}},{\bf{r}} ^{\prime}) = \frac{({\bf{r}} _{\perp }  - {\bf{r}} _{\perp } ^{\prime}) _{j}}{\vert {\bf{r}} _{\perp }  - {\bf{r}} _{\perp } ^{\prime} \vert ^{2} } \, \left( \sqrt{\vert {\bf{r}} _{\perp }  - {\bf{r}} _{\perp } ^{\prime} \vert ^{2} + (\vert z \vert  + z ^{\prime}) ^{2} } - (\vert z \vert  + z ^{\prime}) \right) .
\end{align}




\section{Frequency-dependent Green's function}

\subsection{Derivation of the Green's functions} \label{Derivation_GF_omega}

Substituting the zeroth-order solutions (\ref{Zeroth-order-T_dep}) into Eqs. (\ref{PotPhi-ThetaN}) and (\ref{PotA-ThetaN}) one finds that the first order solution are given by
\begin{align}
\phi ^{(1)} ({\bf{r}} , \omega) &= - \epsilon _{zij} \partial _{i} \int   \mathcal{G} _{\omega} ({\bf{r}} , {\bf{r}}' ) \, J _{j} ({\bf{r}}' , \omega) \,  d^{3} {\bf{r}}'  \label{PotE_App} \\ A ^{(1)} _{i} ({\bf{r}} , \omega) &=  - \epsilon _{izj}  \partial _{j} \int  \mathcal{G}_{\omega} ({\bf{r}} , {\bf{r}}' )   \rho ({\bf{r}}')  d ^{3} {\bf{r}}'  + i \frac{\omega }{v ^{2}}  \epsilon _{izj}  \int \mathcal{G} _{\omega} ({\bf{r}} , {\bf{r}}' ) \,  J _{j} ({\bf{r}}' , \omega)   \, d ^{3} {\bf{r}}' , \label{PotA_App}
\end{align}
where we have defined
\begin{align}
\mathcal{G} _{\omega} ({\bf{r}} , {\bf{r}}' ) = \tilde{\alpha} \frac{ \mu}{\epsilon}  \sqrt{\frac{\epsilon _{0}}{\mu _{0}}} \int G _{\omega} ({\bf{r}},{\bf{r}}'' _{\perp}) \, G _{\omega} ({\bf{r}}'' _{\perp},{\bf{r}}') \, d ^{2} {\bf{r}}'' _{\perp} .  \label{Generic_Int}
\end{align}
We now perform son manipulations in Eqs. (\ref{PotE_App}) and (\ref{PotA_App}). Taking the divergence of Eq. (\ref{PotA_App}) we obtain
\begin{align}
\partial _{i} A ^{(1)} _{i} ({\bf{r}} , \omega) &= i \frac{\omega }{v ^{2}}  \epsilon _{izj} \partial _{i} \int \mathcal{G} _{\omega} ({\bf{r}} , {\bf{r}}' ) \,  J _{j} ({\bf{r}}' , \omega)   \, d ^{3} {\bf{r}}' =  i \frac{\omega }{v ^{2}}  \phi ^{(1)} ({\bf{r}} , \omega) ,
\end{align}
where we used that $\epsilon _{zij} \partial _{i} \partial _{j} = 0$. So, one we have computed the vector potential, the scalar potential becomes fully determined. In order to manipulate Eq. (\ref{PotA_App}) we use the local charge conservation: $\nabla \cdot {\bf{J}} = i \omega \rho$. Substituting the charge density in  Eq. (\ref{PotA_App}) produces
\begin{align}
A ^{(1)} _{i} ({\bf{r}} , \omega) &=  - \frac{1}{i \omega} \epsilon _{izj}  \partial _{j} \int  \mathcal{G}_{\omega} ({\bf{r}} , {\bf{r}}' )  [ \partial' _{k}  J _{k} ({\bf{r}}' , \omega) ]  d ^{3} {\bf{r}}'  + i \frac{\omega }{v ^{2}}  \epsilon _{izj}  \int \mathcal{G} _{\omega} ({\bf{r}} , {\bf{r}}' ) \,  J _{j} ({\bf{r}}' , \omega)   \, d ^{3} {\bf{r}}' . 
\end{align}
Integrating by parts and using the identity $\partial' _{i} \mathcal{G}_{\omega} ({\bf{r}} , {\bf{r}}' )  = - \partial _{i} \mathcal{G}_{\omega} ({\bf{r}} , {\bf{r}}' )$ we  obtain
\begin{align}
A ^{(1)} _{i} ({\bf{r}} , \omega) &=  \frac{1}{i \omega} \epsilon _{zij}  \partial _{j} \partial _{k} \int  \mathcal{G}_{\omega} ({\bf{r}} , {\bf{r}}' )  \,   J _{k} ({\bf{r}}' , \omega) \,  d ^{3} {\bf{r}}'  + i \frac{\omega }{v ^{2}}  \epsilon _{izk}  \int \mathcal{G} _{\omega} ({\bf{r}} , {\bf{r}}' ) \,  J _{k} ({\bf{r}}' , \omega)   \, d ^{3} {\bf{r}}' \notag \\ &= \frac{i}{\omega} \left( -  \epsilon _{zij}  \partial _{j} \partial _{k} +  \frac{\omega ^{2} }{v ^{2}}  \epsilon _{izk} \right)  \int \mathcal{G} _{\omega} ({\bf{r}} , {\bf{r}}' ) \,  J _{k} ({\bf{r}}' , \omega)   \, d ^{3} {\bf{r}}' ,
\end{align}
which can be written as
\begin{align}
A ^{(1)} _{i} ({\bf{r}} , \omega) &=  \int \mathcal{G} _{ij} ({\bf{r}} , {\bf{r}}' ) \,  J _{j} ({\bf{r}}' , \omega)   \, d ^{3} {\bf{r}}' ,
\end{align}
where
\begin{align}
 \mathcal{G} _{ij} ({\bf{r}} , {\bf{r}}' )  &= \frac{i}{\omega} \left( -  \epsilon _{zij}  \partial _{j} \partial _{k} +  \frac{\omega ^{2} }{v ^{2}}  \epsilon _{izk} \right)  \mathcal{G} _{\omega} ({\bf{r}} , {\bf{r}}' ) . 
\end{align}
This expression corresponds to Eq. (\ref{Gij-time}) of the main text.

In the following we derive an exact expression for the function $\mathcal{G} _{\omega} ({\bf{r}} , {\bf{r}}' ) $, which cannot be expressed in term of known functions. We also obtain an analytical expression in the \textit{radiation zone} (far field approximation) by using two different methods.

\subsection{Exact expression for the function $\mathcal{G} _{\omega} ({\bf{r}} , {\bf{r}}' ) $} \label{ExactApp}

The integral we have to evaluate is given by Eq. (\ref{Generic_Int}). Substituting the frequency-dependent Green's function (\ref{Usual_Green_Frequency}) the integral becomes:
\begin{align}
\mathcal{G} _{\omega} ({\bf{r}}, {\bf{r}} ^{\prime} ) = \tilde{\alpha} \frac{ \mu}{\epsilon}  \sqrt{\frac{\epsilon _{0}}{\mu _{0}}} \int d ^{2} {\bf{r}} ^{\prime \prime } _{\perp} \, \frac{e ^{i k _{0}  \vert {\bf{r}} - {\bf{r}} _{\perp} ^{\prime \prime } \vert }}{ 4 \pi \, \vert {\bf{r}} - {\bf{r}} _{\perp} ^{\prime \prime } \vert } \, \frac{e ^{i k _{0} \vert {\bf{r}} _{\perp} ^{\prime \prime } - {\bf{r}} ^{\prime} \vert }}{ 4 \pi \, \vert {\bf{r}} _{\perp} ^{\prime \prime } - {\bf{r}} ^{\prime}  \vert } . \label{Basic_Int}
\end{align}
To evaluate this integral we now  use the Sommerfeld identity:
\begin{align}
\frac{e ^{i k _{0} \vert {\bf{r}} \vert }}{4\pi \, \vert {\bf{r}} \vert  } = \int _{\Sigma _{{\bf{k}} _{\perp}}} \frac{d ^{2} {\bf{k}} _{\perp}}{(2 \pi ) ^{2} } e ^{i {\bf{k}} _{\perp} \cdot {\bf{r}} _{\perp}} \frac{i e ^{ i \sqrt{ k _{0} ^{2} - k _{\perp} ^{2}} \vert z \vert }}{2 \sqrt{ k _{0} ^{2} - k _{\perp} ^{2} }}  , \label{Identity}
\end{align}
where $\Sigma _{{\bf{k}} _{\perp}}$ is the two-dimensional space $k_{x}$-$k_{y}$ and ${\bf{r}} = {\bf{r}} _{\perp} + \vert z \vert \hat{{\bf{e}}} _{z} $. Using this representation twice in Eq. (\ref{Basic_Int}) we get
\begin{align}
\mathcal{G} _{\omega} ({\bf{r}}, {\bf{r}} ^{\prime} ) = \tilde{\alpha} \frac{ \mu}{\epsilon}  \sqrt{\frac{\epsilon _{0}}{\mu _{0}}} \int _{\Sigma _{{\bf{k}} _{\perp}}} \frac{d ^{2} {\bf{k}} _{\perp}}{(2 \pi ) ^{2} }  \int _{\Sigma _{{\bf{k}}' _{\perp}}} \frac{d ^{2} {\bf{k}}' _{\perp}}{(2 \pi ) ^{2} }  \, e ^{i ( {\bf{k}} _{\perp} \cdot  {\bf{r}}  _{\perp}  -  {\bf{k}}' _{\perp} \cdot   {\bf{r}}  _{\perp}  ^{\prime } ) }  \frac{i e ^{ i \sqrt{ k _{0} ^{2} - k _{\perp} ^{2}} \vert z \vert }}{2 \sqrt{ k _{0} ^{2} - k _{\perp} ^{2} }}  \,  \frac{i e ^{ i \sqrt{ k _{0} ^{2} - k _{\perp} ^{\prime \, 2}} \vert z ^{\prime } \vert }}{2 \sqrt{ k _{0} ^{2} - k _{\perp} ^{\prime \, 2} }} \int d ^{2} {\bf{r}} ^{\prime \prime } _{\perp} \,    e ^{i ( {\bf{k}}' _{\perp}  - {\bf{k}} _{\perp} ) \cdot   {\bf{r}} _{\perp} ^{\prime \prime }  }
\end{align}
Integration over $ {\bf{r}} _{\perp} ^{\prime \prime } $ produces a Dirac delta in momentum, which allow us to evaluate one of the momentum integrals. We obtain:
\begin{align}
\mathcal{G} _{\omega} ({\bf{r}}, {\bf{r}} ^{\prime} ) = - \frac{\tilde{\alpha}}{4}  \frac{ \mu}{\epsilon}  \sqrt{\frac{\epsilon _{0}}{\mu _{0}}} \int _{\Sigma _{{\bf{k}} _{\perp}}} \frac{d ^{2} {\bf{k}} _{\perp}}{(2 \pi ) ^{2} }   e ^{i {\bf{k}} _{\perp} \cdot  (  {\bf{r}}  _{\perp}  -  {\bf{r}}  _{\perp}  ^{\prime } ) }  \frac{ e ^{ i \sqrt{ k _{0} ^{2} - k _{\perp} ^{2}} ( \vert z \vert + \vert z ^{\prime } \vert ) }}{  k _{0} ^{2} - k _{\perp} ^{2} }  , 
\end{align}
As before, we now introduce the polar coordinates $(k _{x}, k _{y} ) = k _{\perp} ( \cos \varphi , \sin \varphi ) $ and choose the vector ${\bf{r}} _{\perp }  - {\bf{r}} _{\perp } ^{\prime}$ along the $k _{x}$ direction. So, the angular integral yields
\begin{align}
\mathcal{G} _{\omega} ({\bf{r}}, {\bf{r}} ^{\prime} ) = - \frac{\tilde{\alpha}}{8 \pi} \frac{ \mu}{\epsilon}  \sqrt{\frac{\epsilon _{0}}{\mu _{0}}} \int _{0} ^{\infty} d k _{\perp}  \frac{e ^{i \sqrt{ k _{0} ^{2}  - k _{\perp} ^{2} } ( \vert z \vert + \vert z ^{\prime} \vert )}}{ k _{0} ^{2} - k _{\perp} ^{2} }  k _{\perp} \, J _{0} ( k _{\perp} \vert {\bf{r}} _{\perp} - {\bf{r}} ^{\prime} _{\perp} \vert ) .  \label{Integral}
\end{align}
This is an exact expression for the required function. It corresponds to Eq. (\ref{ExactG_omega}) of the main text.

\subsection{Far-field approximation} \label{Far_field_App}

Here we shall approximate the integral (\ref{Integral}) in the radiation zone, i.e. for $\rho, \vert z\vert  \gg 1$. For crosscheking our result, we compute the integral by two different methods. One of them used the Sommerfeld integral (\textit{Method 1}) and the other is the stationary phase approximation (\textit{Method 2}).\\

\textit{\textbf{Method 1}}.  We start by using the identity $q^{-1} = \int _{0} ^{\infty} d \xi \, e ^{-q \xi }$ in Eq. (\ref{Integral}), for $q = \sqrt{k _{0} ^{2} - k _{\perp} ^{2}}$ to obtain
\begin{align}
 \mathcal{G} _{\omega} ({\bf{r}}, {\bf{r}} ^{\prime} ) &= -  \frac{\tilde{\alpha}}{8 \pi} \frac{ \mu}{\epsilon}  \sqrt{\frac{\epsilon _{0}}{\mu _{0}}}  \int _{0} ^{\infty} d \xi \,  \int _{0} ^{\infty} d k _{\perp} \, \frac{k _{\perp}}{ \sqrt{k _{0} ^{2} - k _{\perp} ^{2}} } e ^{i \sqrt{k _{0} ^{2} - k _{\perp} ^{2}} (   \vert z \vert + \vert z ^{\prime} \vert  + i \xi ) }  \, J _{0} ( k _{\perp} \vert {\bf{r}} _{\perp} - {\bf{r}} ^{\prime} _{\perp} \vert ) .
\end{align}
Using the Sommerfeld integral (\ref{Identity}) this expression can be written in the simple form:
\begin{align}
    \mathcal{G} _{\omega} ({\bf{r}}, {\bf{r}} ^{\prime} ) =  i  \frac{\tilde{\alpha}}{8 \pi} \frac{ \mu}{\epsilon}  \sqrt{\frac{\epsilon _{0}}{\mu _{0}}}  \int _{0} ^{\infty} d \xi \, \frac{ e ^{i k _{0} R _{\xi} }}{ R _{\xi} }  , \label{Integral3}
\end{align}
where $R _{\xi} \equiv R _{\xi} ({\bf{r}}, {\bf{r}} ^{\prime} )= \sqrt{ \vert {\bf{r}} _{\perp} - {\bf{r}} ^{\prime} _{\perp} \vert  ^{2} + ( \vert z \vert + \vert z ^{\prime} \vert + i \xi ) ^{2} }$, such that $R _{0} =  \sqrt{ \vert {\bf{r}} _{\perp} - {\bf{r}} ^{\prime} _{\perp} \vert  ^{2} + ( \vert z \vert + \vert z ^{\prime} \vert  ) ^{2} }$. Next we shall approximate this integral  in the far-field region. To this end we use the fact that $R _{\xi} =  R _{0} \sqrt{1 + \frac{2 i \xi ( \vert z \vert + \vert z ^{\prime} \vert  ) - \xi ^{2}}{ R _{0} ^{2}}  }$ to rewritte the integral (\ref{Integral3}) as
 \begin{align}
    \mathcal{G} _{\omega} ({\bf{r}}, {\bf{r}} ^{\prime} ) = i  \frac{\tilde{\alpha}}{8 \pi} \frac{ \mu}{\epsilon}  \sqrt{\frac{\epsilon _{0}}{\mu _{0}}}    \int _{0} ^{\infty} d \chi \, \frac{ e ^{i \zeta  \sqrt{1 + 2 i \chi \tilde{z} - \chi ^{2}   }   }}{  \sqrt{1 + 2 i \chi \tilde{z} - \chi ^{2} } }  , \label{Integral4}
\end{align}
where $\zeta = k _{0} R _{0}$ and $\tilde{z} =  ( \vert z \vert + \vert z ^{\prime} \vert  )  / R _{0}$. Numerical evaluation of this integral for $\zeta \gg 1$ and $\tilde{z} \ll 1$ reveals that only small values of $\chi$ contribute to the integral. So we can expand the root and retain solely the leading order, i.e.
\begin{align}
    \mathcal{G} _{\omega} ({\bf{r}}, {\bf{r}} ^{\prime} ) \approx  i  \frac{\tilde{\alpha}}{8 \pi} \frac{ \mu}{\epsilon}  \sqrt{\frac{\epsilon _{0}}{\mu _{0}}} e ^{ i \zeta }  \int _{0} ^{\infty} d \chi \,  \frac{e ^{ -  \left(   \zeta \tilde{z}   \right) \chi }}{1 + i \chi \zeta} = i  \frac{\tilde{\alpha}}{8 \pi} \frac{ \mu}{\epsilon}  \sqrt{\frac{\epsilon _{0}}{\mu _{0}}}  \frac{1}{  \tilde{ z}}  \left[ \pi +i \, \text{Ei}(i \zeta ) \right]  ,
\end{align}
whose approximation for $\zeta \gg 1$ is
\begin{align}
\mathcal{G} _{\omega} ({\bf{r}}, {\bf{r}} ^{\prime} ) \approx  i  \frac{\tilde{\alpha}}{8 \pi} \frac{ \mu}{\epsilon}  \sqrt{\frac{\epsilon _{0}}{\mu _{0}}} \frac{e ^{ i k _{0} R _{0} }   }{ k _{0} ( \vert z \vert + \vert z ^{\prime} \vert  ) }  .   \label{Res1}
\end{align}
This result corresponds to Eq. (\ref{Aprox_G_omega}) of the main text.\\

\textit{\textbf{Method 2}}. Since the Bessel function and the complex exponential in the integrand of Eq. (\ref{Integral}) are rapidly oscillating functions in the far-field region, one can use the stationary phase approximation to evaluate the integral.  To this end we first use the identity $J _{0} (x) = \tfrac{1}{2}  [ H ^{(1)} _{0} (x) + H ^{(2)} _{0} (x)  ]$ together with the reflection formula $H ^{(1)} _{0} (e^{i \pi } x) = - H ^{(2)} _{0} (x)$, where $H ^{(1)} _{0}$ and $H ^{(2)} _{0}$ are the Hankel functions.  In this case, the integral (\ref{Integral}) can be written as
\begin{align}
\mathcal{G} _{\omega} ({\bf{r}}, {\bf{r}} ^{\prime} ) &=  - \frac{\tilde{\alpha}}{16 \pi}    \frac{ \mu}{\epsilon}  \sqrt{\frac{\epsilon _{0}}{\mu _{0}}}   \int _{- \infty} ^{\infty} d k _{\perp} \, \frac{k _{\perp}}{k _{0} ^{2} - k _{\perp} ^{2}} e ^{i k _{z} ( \vert z \vert + \vert z ^{\prime} \vert  ) } \,  H ^{(1)} _{0} (k _{\perp} \vert {\bf{r}} _{\perp} - {\bf{r}} ^{\prime} _{\perp} \vert  ) .  \label{Integral5}
\end{align}
We now implement the far-field approximation into the Hankel function. For $x \gg 1$ we have $H ^{(1)} _{0} (x) \approx \sqrt{\frac{2}{\pi x}} e ^{ i x - i \frac{\pi}{4} }$,  and the integral (\ref{Integral5}) becomes
 \begin{align}
\mathcal{G} _{\omega} ({\bf{r}}, {\bf{r}} ^{\prime} ) & \approx  - \frac{\tilde{\alpha}}{16 \pi}    \frac{ \mu}{\epsilon}  \sqrt{\frac{\epsilon _{0}}{\mu _{0}}}   \int _{- \infty} ^{\infty} d k _{\perp} \, \frac{k _{\perp}}{k _{0} ^{2} - k _{\perp} ^{2}}   \sqrt{\frac{2}{\pi k _{\perp} \vert {\bf{r}} _{\perp} - {\bf{r}} ^{\prime} _{\perp} \vert  }} \exp \left[ i \vert {\bf{r}} _{\perp} - {\bf{r}} ^{\prime} _{\perp} \vert  \left( k _{\perp}  +  \sqrt{k _{0} ^{2} - k _{\perp} ^{2}} \frac{\vert z \vert + \vert z ^{\prime} \vert }{\vert {\bf{r}} _{\perp} - {\bf{r}} ^{\prime} _{\perp} \vert}  \right) - i \frac{\pi}{4}  \right]     .  \label{Integral6}
\end{align}
We observe that, in the far-field, the exponential factor is a rapidly oscillating function of $k _{\perp}$  and hence we can apply the stationary phase approximation to evaluate the dominant contribution.  A simple calculation reveals that the stationary phase point locates at
\begin{align}
k _{\perp} ^{\ast} = \frac{k _{0}}{\sqrt{1 + \left( \frac{\vert z \vert + \vert z ^{\prime} \vert }{\vert {\bf{r}} _{\perp} - {\bf{r}} ^{\prime} _{\perp} \vert} \right) ^{2}}} .
\end{align}
Finally, we use the well-known asymptotic formula
\begin{align}
\int _{\mathbb{R}} g (x) e ^{i \lambda f (x)} dx \approx  g (x _{0}) e ^{i \lambda f (x _{0}) + i \frac{\pi}{4} \mbox{sgn}(f'' (x _{0}))} \sqrt{ \frac{2 \pi}{ \lambda \vert f '' (x _{0}) \vert } } ,
\end{align}
where $\lambda \gg 1$ and $f' (x_{0}) = 0$, to obtain
\begin{align}
\mathcal{G} _{\omega} ({\bf{r}}, {\bf{r}} ^{\prime} ) \approx  i  \frac{\tilde{\alpha}}{8 \pi} \frac{ \mu}{\epsilon}  \sqrt{\frac{\epsilon _{0}}{\mu _{0}}} \frac{e ^{ i k _{0} R _{0} }   }{ k _{0} ( \vert z \vert + \vert z ^{\prime} \vert  ) }  ,
\end{align}
which is again the result of Eq. (\ref{Res1}). This validates our expression for the far-field approximation.

\section{Evaluation of the first-order potentials for a short linear antenna} \label{App_Antenna}

Our starting point here is Eq. (\ref{A-Main-time}) for the components of the vector potential. The first order contribution in the frequency-domain is read off as
\begin{align}
A _{i}  ^{(1)} ({\bf{r}} , \omega ) &=  i \frac{1}{\omega} \left[ (\hat{\bf{e}} _{z} \times \nabla ) _{i} \partial _{j}  - k _{0} ^{2} \,  \epsilon _{ijz}   \right]  \int  \mathcal{G} _{\omega}  ({\bf{r}} , {\bf{r}}' )  \, J_{j} ({\bf{r}} ^{\prime} , \omega ) \, d ^{3} {\bf{r}} ^{\prime} ,   
\end{align}
where $\mathcal{G} _{\omega} ({\bf{r}} , {\bf{r}}' )$ is the corresponding Green's function. The potential in the time-domain is computed by inverse Fourier transforming the above expression. We obtain:
\begin{align}
A _{i}  ^{(1)} ({\bf{r}} , t ) &= i \frac{1}{2 \pi} \left[ (\hat{\bf{e}} _{z} \times \nabla ) _{i} \partial _{j}  + \frac{1}{v ^{2}} \frac{\partial ^{2}}{\partial t ^{2}} \,  \epsilon _{ijz}   \right] \int d \omega \, e^{- i \omega t}  \frac{1}{\omega}   \int  \mathcal{G} _{\omega}  ({\bf{r}} , {\bf{r}}' )  \, J_{j} ({\bf{r}} ^{\prime} , \omega ) \, d ^{3} {\bf{r}} ^{\prime}    .  \label{Pot_Time}
\end{align}
The current source (\ref{Current-Antenna}) in the frequency domain is 
\begin{align}
{\bf{J}} ({\bf{r}} ,  \omega ) &= \int e ^{  i \omega t} {\bf{J}} ({\bf{r}} , t ) dt  = I \,  \delta (y) \delta (z-z _{0}) \Theta (d - \vert x \vert ) \, \hat{{\bf{e}}}_{x} \frac{\pi}{i} \left[ \delta (\omega + \omega _{0}) - \delta (\omega - \omega _{0}) \right] .  \label{Current-frequency}
\end{align}
Substituting (\ref{Current-frequency}) into Eq. (\ref{Pot_Time}) and integrating the frequency one obtains
\begin{align}
A _{i}  ^{(1)} ({\bf{r}} , t ) &= -  \frac{1}{2 \pi} \left[ (\hat{\bf{e}} _{z} \times \nabla ) _{i} \partial _{x}  + \frac{1}{v ^{2}} \frac{\partial ^{2}}{\partial t ^{2}} \,  \epsilon _{ixz}   \right]   \frac{I \pi}{\omega _{0}} \int _{-d} ^{+d}  d x ^{\prime} \,  \left[ e^{ i \omega _{0} t} \mathcal{G} _{- \omega _{0}}  ({\bf{r}} , {\bf{r}}'' ) + e^{- i \omega _{0} t} \mathcal{G} _{\omega _{0} }  ({\bf{r}} , {\bf{r}}'' ) \right]  ,
\end{align}
where ${\bf{r}} ^{\prime \prime} = x' \hat{{\bf{e}}} _{x} + z _{0} \hat{{\bf{e}}} _{z} $. Since we are interested in the far-field domain, we use the Green's function (\ref{Aprox_G_omega}). In such case the vector potential becomes
\begin{align}
A _{i}  ^{(1)} ({\bf{r}} , t ) = - \frac{\tilde{\alpha}}{8 \pi } \frac{\mu}{\epsilon} \sqrt{\frac{\epsilon _{0}}{\mu _{0}}} \frac{I}{ \omega _{0} k _{0} (\vert z \vert + z _{0})}  \left[ (\hat{\bf{e}} _{z} \times \nabla ) _{i} \partial _{x}  - \frac{\omega _{0} ^{2}}{v ^{2}} \,  \epsilon _{ixz}   \right]   \int _{-d} ^{+d}  d x ^{\prime}  \sin  \left[  \omega _{0} \left( t - \frac{  \vert {\bf{r}} ^{\ast} - {\bf{r}} ' _{i} \vert  }{v}  \right)   \right] , \label{Full_Vect_Pot}
\end{align}
where ${\bf{r}} ' _{i} (x') = x' \hat{{\bf{e}}}_{x} - z _{0} \hat{{\bf{e}}}_{z} $ is the image point of  ${\bf{r}} ' (x')$ and ${\bf{r}} ^{\ast}  = x \hat{{\bf{e}}}_{x} + y \hat{{\bf{e}}}_{y} + \vert z \vert \hat{{\bf{e}}}_{z} $. Equations (\ref{AX}) and (\ref{AY}) of the main text are derived from this result by taking $i=x$ and $i=y$, respectively. Now, to derive the scalar potential, we appeal to the Lorentz gauge condition. Taking the divergence of the vector potential we obtain
\begin{align}
\partial _{i} A _{i}  ^{(1)} ({\bf{r}} , \omega ) &=  - \frac{\tilde{\alpha}}{8 \pi } \frac{\mu}{\epsilon} \sqrt{\frac{\epsilon _{0}}{\mu _{0}}} \frac{I}{ \omega _{0} k _{0} (\vert z \vert + z _{0})}    \frac{\omega _{0} ^{2}}{v ^{2}} \, \frac{\partial}{\partial y} \int _{-d} ^{+d}  d x ^{\prime}  \sin  \left[  \omega _{0} \left( t - \frac{  \vert {\bf{r}} ^{\ast} - {\bf{r}} ' _{i} \vert  }{v}  \right)   \right] = - \frac{1}{v ^{2}} \frac{\partial}{\partial t} \phi ^{(1)}  ({\bf{r}} ,  t ),
\end{align}
 wherefrom we find
\begin{align}
\phi ^{(1)}  ({\bf{r}} ,  t )  =- \frac{\tilde{\alpha}}{8 \pi } \frac{\mu}{\epsilon} \sqrt{\frac{\epsilon _{0}}{\mu _{0}}} \frac{I}{ k _{0} (\vert z \vert + z _{0})}  \, \frac{\partial}{\partial y} \int _{-d} ^{+d}  d x ^{\prime}  \cos  \left[  \omega _{0} \left( t - \frac{ \vert {\bf{r}} ^{\ast} - {\bf{r}} ' _{0i} \vert  }{v}  \right)   \right]  ,
\end{align}
which, after taking the $y$-derivative, corresponds to Eq. (\ref{Scalar_Pot}) of the main text.

\section{Derivation of Eq. (\ref{Radiated_Power})}

The angular distribution of the radiated power is given by $\frac{dP}{d \Omega} =   \frac{v}{\mu}  r ^{2}   \vert {\bf{B}} \vert ^{2} $. Keeping linear order in the magnetic field, i.e. $\vert {\bf{B}} \vert ^{2}  \approx   \vert {\bf{B}} ^{(0)} \vert ^{2} + 2  {\bf{B}}  ^{(0)}  \cdot {\bf{B}}  ^{(1)}  $, one finds $\frac{dP}{d \Omega}   \approx   \frac{v}{\mu}  r ^{2}  \left[ \vert {\bf{B}} ^{(0)} \vert ^{2} + 2  {\bf{B}}  ^{(0)}  \cdot {\bf{B}}  ^{(1)} \right]$.  Substituting the zeroth (\ref{B0}) and first (\ref{B1}) order magnetic fields we obtain
\begin{align}
\frac{dP}{d \Omega} & \approx   \frac{v}{\mu}  \left( \frac{\mu  I _{0} }{2 \pi }  \right) ^{2}  \cos \left[ \omega _{0} \left( t -  \frac{ r ^{2} - z _{0} z }{v r}  \right) \right]  \sin ^{2} \left( \frac{ \omega _{0} d    }{v}  \frac{x}{r} \right)   \left\lbrace  \cos  \left[ \omega _{0} \left( t -  \frac{ r ^{2} - z _{0} z }{v r}  \right) \right]    \, \frac{1  - (x/r) ^{2}}{(x/r) ^{2}} \right. \notag \\ & \hspace{7cm} \left.  + 2 \tilde{\alpha}    \sqrt{\frac{\epsilon _{0}}{\mu _{0}}  \frac{\mu}{\epsilon} }    \cos  \left[ \omega _{0} \left( t -  \frac{r ^{2 } +  z _{0} \vert z \vert  }{v \, r  }   \right)    \right] \,  \frac{  y/x }{  \vert z \vert + z _{0} }    r  \right\rbrace ,
\end{align}
where we used that $( \hat{{\bf{e}}} _{x} \times \hat{{\bf{n}}} )  \cdot ( \hat{{\bf{e}}} _{x} \times \hat{{\bf{n}}} )  = 1  - ( x/r ) ^{2} $ and $( \hat{{\bf{e}}} _{x} \times \hat{{\bf{n}}} )  \cdot ( \hat{{\bf{e}}} _{y} \times \hat{{\bf{n}}} ) = - xy/r ^{2}$. 

Finally, averaging in time over a complete cycle yields
\begin{align}
\left< \frac{dP}{d \Omega} \right> & \approx   \frac{v}{2 \mu}  \left( \frac{\mu  I _{0} }{2 \pi }  \right) ^{2}   \sin ^{2} \left( \frac{ \omega _{0} d    }{v}  \frac{x}{r} \right)   \left\lbrace    \frac{1  - (x/r) ^{2}}{(x/r) ^{2}}    + 2 \tilde{\alpha}    \sqrt{\frac{\epsilon _{0}}{\mu _{0}}  \frac{\mu}{\epsilon} }    \cos \left[ \frac{ ( z + \vert z \vert )  z _{0} \omega  }{r v } \right]  \,  \frac{  y/x }{  \vert z \vert + z _{0} }    r  \right\rbrace ,
\end{align}
where we have used:
\begin{align}
\lim _{T \to \infty} \frac{1}{2 T} \int _{-T} ^{+T} \cos ^{2} \left[ \omega _{0} \left( t -  \frac{ r ^{2} - z _{0} z }{v r}  \right) \right]  dt & = \frac{1}{2} , \\ \lim _{T \to \infty} \frac{1}{2 T} \int _{-T} ^{+T} \cos  \left[ \omega _{0} \left( t -  \frac{ r ^{2} - z _{0} z }{v r}  \right) \right] \cos  \left[ \omega _{0} \left( t -  \frac{r ^{2 } +  z _{0} \vert z \vert  }{v r  }   \right)    \right] dt & = \frac{1}{2} \cos \left[ \frac{ ( z + \vert z \vert )  z _{0} \omega  }{r v } \right] .
\end{align}
These results stablish Eq. (\ref{Ang_Dist_Rad_Power}).

\bibliography{references.bib}

\end{document}